\documentclass[prl,aps,amssymb,twocolumn,superscriptaddress, notitlepage, longbibliography]{revtex4-2}

\usepackage{bm,amssymb,amsmath,dsfont,mathrsfs,amstext,latexsym,physics,relsize}
\usepackage{graphicx}
\usepackage[usenames,dvipsnames]{xcolor}
\usepackage[colorlinks=true,citecolor=blue,urlcolor=blue,linkcolor=blue]{hyperref}

\usepackage{orcidlink}
\usepackage[normalem]{ulem}

\usepackage{mathtools}

\makeatletter
\newcommand*{\addFileDependency}[1]{%
  \typeout{(#1)}%
  \@addtofilelist{#1}%
  \IfFileExists{#1}{}{\typeout{No file #1.}}%
}
\makeatother

\begin{document}

\title{Quantum Transition Rates in Arbitrary Physical Processes}
%in 

\author{Adolfo~del Campo\orcidlink{0000-0003-2219-2851}}    
\email{adolfo.delcampo@uni.lu}
\affiliation{Department  of  Physics  and  Materials  Science,  University  of  Luxembourg,  L-1511  Luxembourg, Luxembourg}
\affiliation{Donostia International Physics Center,  E-20018 San Sebasti\'an, Spain}

\author{Andr\'as Grabarits \orcidlink{0000-0002-0633-7195}}
\email{andras.grabarits@uni.lu}
\affiliation{Department  of  Physics  and  Materials  Science,  University  of  Luxembourg,  L-1511  Luxembourg, Luxembourg}

\author{Dmitrii E. Makarov\orcidlink{0000-0002-8421-1846}}
\affiliation{Department of Chemistry, University of Texas at Austin, Austin, Texas 78712, USA}
\affiliation{Oden Institute for Computational Engineering and Sciences,
University of Texas at Austin, Austin, Texas 78712, USA}

\author{Seong-Ho Shinn  \orcidlink{0000-0002-2041-5292}}
\email{seongho.shin@uni.lu}
\affiliation{Department of Physics and Materials Science, University of Luxembourg, L-1511 Luxembourg, Luxembourg}

\begin{abstract}
We introduce a framework for computing time-dependent quantum transition rates (QTRs) that describe the pace of evolution of a quantum state from a given subspace to a target subspace. QTRs are expressed in terms of flux-flux correlators and are shown to obey two complementary quantum speed limits. Our framework readily accommodates the generalization of Hamiltonian dynamics to arbitrary open quantum evolution, 
including quantum measurements. We illustrate how QTRs can be controlled by counterdiabatic driving. 
\end{abstract}

\maketitle

Quantum speed limits (QSLs) are widely used to estimate the minimum time for a process to unfold \cite{Mandelstam45,Margolus98,Deffner2017,Gong22}. More precisely, they determine the minimum time for the time-evolving state to travel a given distance in state space, measured from the initial state. However, QSLs do not account for the direction in state space. A given evolution may leave the initial subspace at a given rate, but may travel in the `wrong' direction for achieving a given task. As such, QSLs are not suitable for determining the completion of a specific process.  In fact, while they can be saturated for specific evolutions \cite{Margolus98,Giovannetti2003,Levitin2009,Campaioli2018,Campaioli2019}, QSLs are often too conservative. This is apparent when estimating the timescale of physical processes, such as thermalization, where estimates based on QSLs can be off by orders of magnitude from the experimentally observed timescales \cite{Gogolin16,Gong22}.

To avoid this limitation, it is necessary to identify the target subspace of the evolution for a specific process to be completed. Let us define a process as a dynamics in which the quantum state of the system moves from a given subspace $A$ to a different target subspace $B$. This involves conditioning the evolution so that the initial quantum state is found in $A$ and the final state is in $B$. This conditioning, lacking in the formulation of QSLs, is implicit in the definition of transition probabilities \cite{Feynman65}, a fundamental object in nonequilibrium phenomena with broad applications from the foundations of physics \cite{TQM1,TQM2} to chemistry, biophysics \cite{Makarov2015}, and quantum technologies.

The notion of a directed quantum evolution towards a target subspace is essential in chemical reaction rate theory \cite{Yamamoto60,Levine1969,McLafferty74,Miller74,Miller83,Miller93,Miller03,Hanggi90,Lidar99,Schatz02}. 
A particularly relevant formulation is that of Yamamoto \cite{Yamamoto60} and Miller \cite{Miller74,Miller83,Miller93,Miller03}.
The focus in chemical reaction theory involves the dynamics along a reaction coordinate. For illustration, consider a collinear chemical reaction that admits a one-dimensional reaction coordinate $\hat{s}$ with conjugate momentum $\hat{p}$. The Hamiltonian is then $\hat{H}=\hat{p}^2/2m+\hat{V}(s)$, that of a particle moving in one dimension along $s$ in the presence of a potential barrier $\hat{V}(s)$.  The subspace associated with the projector $\hat{\Pi}_A=\Theta(-\hat{s})$ is associated with the reactants, and $\hat{\Pi}_B=\Theta(\hat{s})$ is associated with its products. The reaction rate is given by the expression
\begin{equation}
k=\frac{1}{Z(\beta)}\lim_{t\rightarrow\infty}\tr\left[\hat{J}e^{-(\beta/2-it/\hbar)\hat{H}}\Theta(\hat{s})e^{-(\beta/2+it/\hbar)\hat{H}}\right],
\end{equation}
where $Z(\beta)$ is the partition function of the reactants and the quantum flux takes the conventional expression $\hat{J}=\{\hat{p},\delta(\hat{s})\}/(2m)$.
In general, computing this expression is challenging \cite{Lidar99}.
Yet, it can be recast in terms of an integral of the flux-flux correlator, 
\begin{equation}
C_{JJ}(t)=\frac{1}{Z(\beta)}\tr\left[\hat{J}e^{-(\beta/2-it/\hbar)\hat{H}}\hat{J}e^{-(\beta/2+it/\hbar)\hat{H}}\right],
\end{equation}
as
\begin{equation}
k=\int_0^tdt^\prime C_{JJ}\left(t^\prime\right).
\end{equation}
This relation \cite{Miller83}, reminiscent of the linear response theory \cite{Yamamoto60}, has spurred substantial progress in the calculation of chemical reaction rates \cite{Thompson1997,Chakraborty2005,Venkataraman2007,BoseNancy17}.

Despite its success, we notice the following limitations inherent to such formulation: (i)
It relies on the long-time asymptotic and scattering theory. As such, it is unsuitable to capture the short and intermediate time dependence of the pace of the underlying physical process \cite{Fonda78,Beau17b,BoseNancy17,Ness21}, which is particularly relevant for quantum control \cite{Rice2000}. (ii) It is restricted to Hamiltonian evolution. A generalization to arbitrary dynamics that includes dissipation and decoherence remains to be elaborated, despite related progress \cite{Christie2024}. The current formulation is not suitable for accommodating, for example, non-Markovianity associated with memory effects \cite{Rivas14} and quantum measurements \cite{Jordan24}.   (iii) It is designed for chemical reactions characterized by a well-defined reaction coordinate, in terms of which the subspaces of reactants and products are identified, with the condition $\hat{\Pi}_A+\hat{\Pi}_B=\mathbb{I}$.

In this work, we formulate a theory of quantum transition dynamics between two subspaces $A$ and $B$ of the Hilbert space of the system. Our formulation applies to arbitrary quantum evolution and readily accommodates the inclusion of quantum measurements. Furthermore, it enables us to identify fundamental quantum bounds on the time scale to complete the process. Such bounds provide a natural generalization of QSLs while conditioning the evolution towards a target subspace.

\begin{figure}
    \centering
    \includegraphics[width=1\columnwidth]{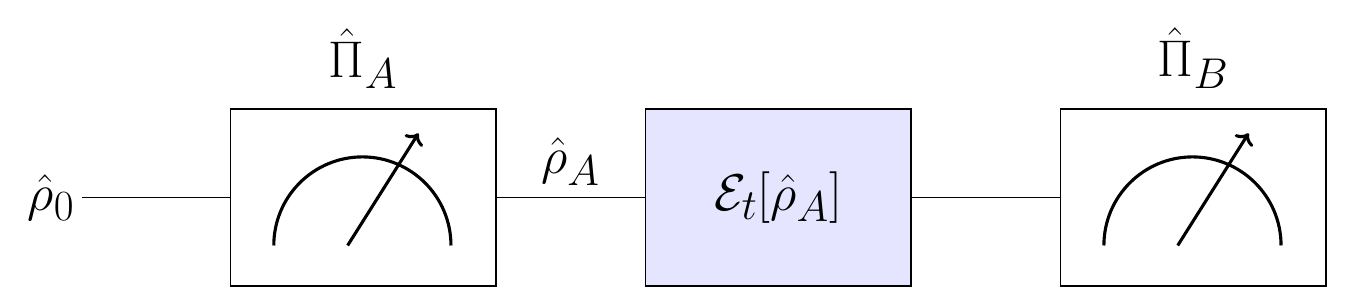}
    \caption{QTRs refine QSLs by using the conditional probability $P(B,t|A,0)$. An initial state $\hat{\rho}_0$ is found in the subspace $\mathcal{H}_A$ by means of a projective measurement of $\hat\Pi_A$. Time evolution is described by a quantum channel $\mathcal{E}_t$ and a second projective measurement, of $\hat\Pi_B$, determines the probability that the state reaches the target subspace $\mathcal{H}_B$. }\label{fig:schematic}
\end{figure}

{\it Quantum Transition Rates (QTRs).}---
Consider a quantum mechanical system described by a Hilbert space $\mathcal{H}$ with two subspaces of interest, $\mathcal{H}_A$ and $\mathcal{H}_B$, associated with projectors $\hat{\Pi}_A$ and $\hat{\Pi}_B$, respectively. 
We consider the general case in which the resolution of the identity reads $\sum_{X=A,B,C,\dots}\hat{\Pi}_X=\mathbb{I}$,  with $\hat{\Pi}_X\hat{\Pi}_Y=\hat{\Pi}_X\delta_{XY}$.
Said differently,  $\mathcal{H}_B$ need not be the orthogonal complement of $\mathcal{H}_A$.
We focus on the transition rate from $\mathcal{H}_A$ to $\mathcal{H}_B$, which we denote by $k_{A\rightarrow B}$.
To this end, we start with the classical definition of a rate
\begin{equation}
k_{A\rightarrow B}(t)=\frac{d}{dt}P(B,t|A),    
\end{equation}
where $P(B,t|A)=P(B,t|A,0)$ is the probability of finding the system at time $t$ in $\mathcal{H}_B$ provided that it was at time $t=0$ in $\mathcal{H}_A$.
We first consider unitary dynamics generated by a possibly time-dependent Hermitian Hamiltonian
$\hat{H}_S \left( t \right)$ in the Schr\"odinger picture. 
The corresponding time evolution operator from $t=0$ to time $t$ is denoted by
$\hat{U}=\mathcal{T}\exp\left(-i\int_0^td\tau\hat{H}_S(\tau)/\hbar\right)$, where $\mathcal{T}$ is the time ordering operator.  
From now on, we will use the Heisenberg picture and write the associated Hamiltonian as
%in this picture as
$
\hat{H} \left( t \right) 
\coloneqq 
\hat{U}^{\dagger} 
\hat{H}_S \left( t \right) 
\hat{U}
$. 
Consider the quantum circuit in Fig. \ref{fig:schematic}. 
Given an initial quantum state $\hat{\rho}_0$, the quantum expression for $P(B,t|A)$ reads
\begin{eqnarray}
P(B,t|A)&=&\frac{1}{P_A}\tr[\hat{U}\hat{\Pi}_A\hat{\rho}_0\hat{\Pi}_A\hat{U}^\dag \hat{\Pi}_B],
\end{eqnarray}
where $P_A=\tr(\hat{\rho}_0\hat{\Pi}_A)$. $P(B,t|A)$ can be measured via the two-time projective measurement scheme \cite{An2015,Dorner_PRL_2013}. Alternatively, note that  $P(B,t|A)=\tr[\hat{\rho}_A \hat{\Pi}_B(t)]$,  upon defining the projection of the state $\hat{\rho}_0$ into the subspace $A$ as $\hat{\rho}_A=\hat{\Pi}_A\hat{\rho}_0\hat{\Pi}_A/P_A$, 
with $\tr\hat{\rho}_A =1$. Further, the projection operator in the Heisenberg picture is given by 
%,  and 
$\hat{\Pi}_B(t)=\hat{U}^\dag \hat{\Pi}_B\hat{U}$.
It follows that
\begin{equation}
k_{A\rightarrow B}(t)=\tr[\hat{\rho}_A \dot{\hat{\Pi}}_B(t)]=\frac{1}{i\hbar}\tr\left(\hat{\rho}_A [\hat{\Pi}_B(t),\hat{H}(t)]\right),    
\end{equation}
where the second equality holds under unitary dynamics generated by a Hamiltonian $\hat{H}(t)$. When the two projectors resolve the identity $\hat{\Pi}_A+\hat{\Pi}_B=\mathbb{I}$, it follows that
\begin{equation}
k_{A\rightarrow B}(t)=-\tr[\hat{\rho}_A \dot{\hat{\Pi}}_A(t)]
=\frac{i}{\hbar}\tr\left(\hat{\rho}_A [\hat{\Pi}_A(t),\hat{H}(t)]\right).    
\end{equation}

In chemical reaction rate theory, it is customary to consider a time-independent Hamiltonian $\hat{H}=\hat{p}^2/2m+\hat{V}(s)$ and complementary projectors $\hat{\Pi}_A=\Theta(-\hat{s})$ and $\hat{\Pi}_B=\Theta(\hat{s})$, whence it follows that
$k_{A\rightarrow B}(t)=\tr[\hat{\rho}_A\hat{J}_B(t)]=\frac{1}{P_A}\tr[\hat{\Pi}_A\hat{\rho}_0\hat{\Pi}_A \hat{J}(t)]$,    
with the quantum flux operator $\hat{J}_B=\hat{J}=\frac{1}{2m}\{\hat{p},\delta(\hat{s})\}$.
Note that the symmetrized flux naturally arises as $\hat{J}=\frac{1}{i\hbar}[\Theta(\hat{s}),\hat{p}^2/2m+\hat{V}(s)]=\frac{1}{i\hbar}[\hat{\Pi}_B,\hat{H}]$.
By analogy, it proves convenient to define an operator for the generalized flux into the subspace $\mathcal{H}_B$ as
\begin{equation}
\label{GenFlux}
\hat{J}_B(t)=\frac{1}{i\hbar}[\hat{\Pi}_B(t),\hat{H}(t)]=:\frac{d}{dt}\hat{\Pi}_B(t),
\end{equation}
where the last equality follows from the Heisenberg equation of motion.
As $\hat{H}(t)$ and $\hat{\Pi}_B(t)$ are Hermitian, $\hat{J}_B(t)$ represents a canonical observable with real eigenvalues.
In terms of it, the QTR is given by
\begin{equation}
k_{A\rightarrow B}(t)=\tr[\hat{\rho}_A \hat{J}_B(t)],
%=\tr[\hat{\rho}_A(t) \hat{J}_B],    
\label{defkAB}
\end{equation}
as the expectation value of the flux operator $\hat{J}_B(t)$ on the initial quantum state within the subspace $\mathcal{H}_A$. We note that for a time-dependent Hamiltonian $\hat{H}(t)$,  $\hat{J}_B(t)$ generally inherits an intrinsic time dependence as
%in the sense that its equation of motion is given by 
%\begin{equation}
    $\frac{d}{dt}\hat{J}_B(t)=\frac{1}{i\hbar}[\hat{J}_B(t),\hat{H}(t)]+\frac{1}{i\hbar}[\hat{\Pi}_B \left( t \right),\dot{\hat{H}} \left( t \right)]$.
%    \label{QTRJ}
%\end{equation}

{\it QTRs from flux-flux correlators.}--- 
One advantage of chemical reaction theory is its ability to determine reaction rates using the flux-flux correlation function, offering both computational efficiency and physical insight \cite{Makarov2015,Miller83,Schatz02}. This result can be extended to QTRs.
Consider the rate of variation of $k_{A\rightarrow B}(t)$ given by $ \frac{d}{dt}k_{A\rightarrow B}(t)=\tr[\hat{\rho}_A\frac{d}{dt}\hat{J}_B(t)]$. Its explicit computation, upon integration, yields
\begin{eqnarray}\label{eq:k_fluxflux}
    k_{A\rightarrow B}(t)&=&-2\int_{0}^t dt'
    %{\rm Re}
    \tr[\hat{J}_B(t')
    \hat{J}_{A, t} 
    %\hat{J}_A
    \hat{\rho}_A]
    \nonumber\\
    & & 
    - 
    \frac{1}{i \hbar} \int_{0}^t d t' 
    \tr \left\lbrack 
    \hat{J}_B \left( t' \right) 
    \left\lbrack 
    \left\lbrack 
    \hat{H} \left( t' \right) 
    , 
    \hat{\rho}_A 
    \right\rbrack 
    , 
    \hat{\Pi}_A 
    \right\rbrack 
    \right\rbrack 
    \nonumber\\
    & & +
    \frac{1}{i\hbar}\int_{0}^tdt'
    \tr[\hat{J}_B(t')\hat{\Pi}_A[\hat{H}(t'),\hat{\rho}_A]\hat{\Pi}_A]
    \nonumber\\
    & & +
    \frac{1}{i\hbar}\int_{0}^tdt'
    \tr[\hat{\rho}_A[\hat{\Pi}_B(t'),\dot{\hat{H}}(t')]]
    ,
\end{eqnarray}
where
$
\hat{J}_{A,t}
\coloneqq  
\left\lbrack 
\hat{\Pi}_A, \hat{H} \left( t \right) 
\right\rbrack / i \hbar 
$. 
%Note that this modified flux operator $\hat{J}_{A,t}$  differs from $\hat{J}_{A} (t=0)$ due to the explicit time-dependent Hamiltonian. 
When $\hat{\Pi}_A=\mathbb{I}-\hat{\Pi}_B$, it follows that
$\hat{J}_{A,t} = -\hat{J}_{B,t}$.
Furthermore, when $[\hat{\rho}_A,\hat{H}_S(t)]=0$, Eq.~\eqref{eq:k_fluxflux} reduces to
\begin{eqnarray}
    k_{A\rightarrow B}(t)=2\int_0^t dt'
    %{\rm Re}
    \tr[\hat{\rho}_A \hat{J}_B(t')\hat{J}_{B,t}].
    \label{fluxfluxeq}
\end{eqnarray}
Thus, the QTR is obtained as the integral of the generalized flux-flux autocorrelation $C_{J_BJ_B}(t)=2\tr[\hat{\rho}_A \hat{J}_B(t')\hat{J}_{B,t}]$, which is reminiscent of the known result in chemical reaction dynamics for time-independent Hamiltonians \cite{Miller83}.
The generalized flux-flux relation (\ref{fluxfluxeq}) relies on the fact that
the flux-flux correlator vanishes at $t=0$, i.e., $C_{J_BJ_B}(t) \left( 0 \right) =0$.

Similarly, for a time-independent Hamiltonian $\hat H_S(t)=\hat H(t)=\hat H$, the last term in Eq.~\eqref{eq:k_fluxflux} trivially vanishes. At $t=0$, $\hat J_{A,t}\rightarrow \hat J_A\coloneqq \hat J_A(0)=\left[\hat\Pi_A,\hat H\right]/(i\hbar)$ and provided that $\left[\hat\rho_A,\hat H\right]=0,\,\hat\Pi_A=\mathbb I-\hat\Pi_B$, the QTR simplifies to
\begin{eqnarray}
    k_{A\rightarrow B}(t)=2\int_0^t dt'
    %{\rm Re}
    \tr[\hat{\rho}_A \hat{J}_B(t')\hat{J}_B].
    \label{fluxfluxeq_timeindep}
\end{eqnarray}

{\it Comparison and generalizations.---}
Our approach offers several advantages over the standard formulation of the chemical reaction theory \cite{Levine1969,Miller74,Miller83,Miller93,Miller03}. First, it applies to arbitrary subspaces $\mathcal{H}_A$ and $\mathcal{H}_B$ of the Hilbert space of the system, without requiring them to be complementary or defined through projectors on a reaction coordinate. Thus, QTRs are applicable beyond the dynamics of chemical reactions and encompass systems with discrete and continuous variables and generic subspaces. By way of example, the computation of $P(B,t|A)$ and QTR  is provided for the Bixon-Jortner model \cite{BixonJortner_1968} for different choices of $\hat{\Pi}_B$ in \cite{SM}. 
Second, our formalism does not rely on scattering theory, and, as the descriptions of non-equilibrium chemical reaction rates \cite{MakarovMetiu1997,BoseNancy17}, 
it applies to any finite time $t$, without involving infinite-time asymptotics. 
As a result, QTRs are generally time-dependent.
In addition, our approach is grounded on the probabilistic interpretation of QTRs and quantum measurement theory. The von Neumann postulate states that the post-measurement state following a projective quantum measurement is given by $\hat{\rho}_A$, justifying the definition of the QTR in (\ref{defkAB}) without ad hoc regularizations.
In doing so, the formulation is closer in spirit to that of QSLs \cite{Deffner2017}. In addition, it is not restricted to unitary evolution or a specific dissipation mechanism \cite{Hanggi90} and can be generalized to arbitrary quantum dynamics, including driving and control, dissipation and decoherence, and quantum measurements. 

Consider a quantum channel $\mathcal{E}_t$ that implements a completely positive trace-preserving (CPTP) map describing the evolution of any initial quantum state $\hat{\rho}_0$ such that $\hat{\rho}_t=\mathcal{E}_t(\hat{\rho}_0)$.
We define the quantum flux into the subspace $\mathcal H_B$ in terms of the adjoint quantum channel $\mathcal{E}_t^\dag$ as 
$\hat{J}_B \left( t \right) 
=
\dot{\mathcal{E}}_t^\dag [
\hat{\Pi}_B 
]$, 
whence it follows that the rate takes the form $k_{A\rightarrow B}(t)=\tr[\hat{\rho}_A \hat{J}_B(t)]=\tr\left(\hat{\rho}_A\dot{\mathcal{E}}_t^\dag [
\hat{\Pi}_B 
]\right)$.

As an example, let us combine unitary time evolution with projective measurements associated with a projector $\hat{\Pi}$. For measurements at regular time intervals $\delta t=t/n$,  the evolution is governed by the dynamical map
$U_n(t)=[\hat{\Pi} \hat{U}(t/n)\hat{\Pi} ]^n$.
For a time-independent Hamiltonian $\hat{H}(t)=\hat H$, it is known that
$\lim_{n\rightarrow\infty} U_n(t)=\exp(-it \hat{\Pi} \hat{H}\hat{\Pi}/\hbar)$ \cite{Facchi08}. It follows that when $\hat{\Pi}$ is chosen as $\hat{\Pi}_B$, see~\cite{SM}, the QTR $k_{A\rightarrow B}$ vanishes identically as a result of the quantum Zeno effect.

For finite time intervals $\delta t=t/n$, Taylor expanding  $\hat{\Pi}_B(t)=U(t,0)^\dag \hat{\Pi}_B(0)\hat{U}(t,0)$ shows that $P(B,t\vert A)=\tr[\hat{\rho}_A\hat{H}\hat{\Pi}_B \hat{H}]t^2/\hbar^2+\mathcal{O}(t^3)$.
Thus, sequential projective measurement of $\hat{\Pi}_B$ at intervals $\delta t$ satisfying
$\delta t\ll \hbar/\tr[\hat{\rho}_A\hat{H}\hat{\Pi}_B \hat{H}]^{1/2}$ suppress the QTR, as detailed in \cite{SM}. 

This result generalizes to driven quantum systems. 
Using the Dyson series for the time evolution operator generated by $\hat{H}(t)$,
one finds, neglecting $\mathcal{O}(t^3)$ corrections,  that
\begin{eqnarray}\label{eq:Zeno_PAB}
  P(B,t\vert A)&&\simeq\frac{1}{\hbar^2}\int_0^t\int_0^t dt_1dt_2\tr[\hat{\rho}_A\hat{H}(t_1)\hat{\Pi}_B \hat{H}(t_2)],\quad\\
  k_{A\rightarrow B}(t)&&\simeq\frac{1}{\hbar^2}\int_0^tdt_1\tr[\hat{\rho}_A\left(\hat{H}(t_1)\hat{\Pi}_B \hat{H}(t)+h.c.\right)].\nonumber
\end{eqnarray}

This generalizes the condition on $\delta t$ in the case of $\hat{H}(t)$ to be smaller than the inverse of the square root of the term involving the double integral. Thus, projective measurements at intervals satisfying $P(B,\delta  t|A)^{1/2}\ll 1$ can be used to suppress QTRs.

{\it Quantum Speed Limits to QTRs.}--- 
The QTR in Eq.~(\ref{defkAB}) is defined as the expectation value of the flux operator $\hat{J}_B$ in the time-dependent state $\hat{\rho}_A(t)$, enabling bounds both on the QTR itself and its rate of change. A Mandelstam-Tamm (MT) type bound can be obtained using the Robertson inequality for mixed states \cite{Gong22}, which implies that the QTR is upper bounded in terms of the conditioned state $\hat{\rho}_A$,
\begin{equation}
\label{absk}
    |k_{A\rightarrow B}(t)|
    \leq \frac{2}{\hbar}\Delta_{\hat{\rho}_A} \hat H(t)
    \sqrt{
    P(B,t\vert A)
    \left[
    1 - 
    P(B,t\vert A) 
    \right]
    },
\end{equation}
where the variance of the observable $\hat{O}$ in the state $\hat{\rho}$ reads $\Delta_{\hat{\rho}}^2 \hat{O}=\tr(\hat{\rho} \hat{O}^2)-\tr(\hat{\rho} \hat{O})^2$, and the property $\tr[\hat\rho_A\hat\Pi^2_B(t)]=\tr[\hat\rho_A\hat\Pi_B(t)]=P(B,t\vert A)$ has been used. Thus, the QTRs are bounded by the energy and the Bernoulli fluctuations of the transition probability.

More generally, QSLs can be formulated in terms of quantum information geometry \cite{Amari2000,Bengtsson2017}. For a trace-class projector $\hat{\Pi}_B$, we introduce the normalized target state $\hat{\pi}_B=\hat{\Pi}_B/d_B$ with $d_B=\tr(\hat{\Pi}_B)$. The distance between the time-evolving state $\hat{\rho}_A(t)=\hat{U}\hat{\rho}_A\hat{U}^{\dag}$ and the target state can be measured by the Bures length 
$\mathcal{L}(\hat{\rho}_A(t),\hat{\pi}_B)=\cos^{-1}\sqrt{F(\hat{\rho}_A(t),\hat{\pi}_B)},$
where $F(\hat{\rho}_A(t),\hat{\pi}_B)=\tr[\sqrt{\sqrt{\hat\rho_A(t)}\hat\pi_B\sqrt{\hat\rho_A(t)}}]^2$ is the Uhlmann fidelity \cite{Uhlmann1992}. Using the super-fidelity bounds \cite{Liang_FidelityBound_2002,Miszczak2008SubAS}, 
\begin{equation}
F(\hat{\rho}_A(t),\hat{\pi}_B)\leq P(B,t|A)d^{-1}_B + M_{\hat\rho_A(t)} M_{\hat\pi_B},
\end{equation}
where $M_{\hat\rho_A(t)}=\sqrt{1-\tr(\hat\rho_A^2(t))}$ and $M_{\hat\pi_B}=\sqrt{1-d^{-1}_B}$ are the mixedness of the density operators. 
As a result, the minimal time to sweep this distance can be lower bounded using the transition probability instead of the Uhlmann fidelity, as $\cos^{-1}\sqrt{P(B,\tau|A)d^{-1}_B+M_{\hat\rho_A(\tau)}\sqrt{1-d^{-1}_B}}\leq \cos^{-1}\sqrt{F(\hat\rho_A(t),\hat\pi_B)}$,
\begin{equation}\label{eq: QSLQTR}
\tau\geq \frac{\hbar \,\cos^{-1}\sqrt{P(B,\tau|A)d^{-1}_B+M_{\hat\rho_A(\tau)}\sqrt{1-d^{-1}_B}}}{\frac{1}{\tau}\int_0^\tau \Delta_{\hat{\rho}_A} \hat H(t) dt}.
\end{equation}
In the general case, this inequality captures the competition between the QTR and MT bounds for arbitrary mixed initial and target subspaces, where $\tau_\mathrm{MT}=\cos^{-1}\sqrt{P(A,t\vert A)}/\left(1/\tau\int_0^\tau\Delta_{\hat\rho_A}\hat H(t) dt\right)$. Explicitly, the QTR-based bound is tighter than the MT bound $\tau_\mathrm{MT}\leq \tau_\mathrm{QTR}$, when the following condition holds
\begin{eqnarray}\label{eq: P_AB_P_AA}
P(B,t|A) + d_B M_{\hat\rho_A(t)}\sqrt{1-d^{-1}_B} \leq d_B P(A,t|A),\nonumber
\end{eqnarray}
where $P(A,t\vert A)$ is the initial state probability.
It reduces to an even less restrictive condition, $P(B,t\vert A)\leq d_BP(A,t\vert A)$, when either $\hat{\rho}_A(t)$ or $\hat{\pi}_B$ is pure, where
$\mathcal{L}(\tau) = \cos^{-1}\sqrt{P(B,\tau|A)d^{-1}_B}$~\cite{Mandelstam45}.
In realistic scenarios, the ranks of the initial and final subspaces are often large, resulting in a tighter QTR-based bound, as also exemplified in the transverse field Ising model (TFIM), the particle-in-a-box with Dirac delta, and in the Bixon-Jortner model in~\cite{SM}.

A complementary QSL results from applying the Mandelstam-Tamm time-energy uncertainty relation to the rate of change of the QTR. 
Using the Cauchy-Schwarz inequality, it follows that
\begin{eqnarray}\hbar\left|\dot{k}_{A\rightarrow B}(t)\right|\le
\Delta_{\hat\rho_A}\hat J_B(t)\Delta_{\hat\rho_A}\hat H(t)+\Delta_{\hat\rho_A}\hat\Pi_B(t)\Delta_{\hat\rho_A}\dot{\hat H}(t),\nonumber\\
\end{eqnarray}
which holds for an arbitrary time-dependent Hamiltonian. 
For completeness, we note that in the time-independent case, the rate of change of the QTR is solely upper-bounded by the product of the energy fluctuations times the flux fluctuations.
This equation can then be rewritten as a time-energy uncertainty relation \cite{Mandelstam45}. To this end, one can introduce the characteristic time scale in which the QTR varies by a factor comparable to $\Delta_{\hat{\rho}_A}\hat J_B(t)$, that is,
$\tau_{\rm QTR}:=\Delta_{\hat{\rho}_A}\hat J_B(t)/\left|\dot{k}_{A\rightarrow B}\right|$,
whence it follows the Mandelstam-Tamm inequality,
    $\tau_{\rm QTR}\,\Delta_{\hat{\rho}_A} \hat H(t)\geq\frac{\hbar}{2}.$

{\it Example: QTRs under Counterdiabatic Driving.} 
Counterdiabatic driving (CD) is a quantum control technique that allows speeding up or slowing down a reference quantum evolution by means of an auxiliary counterdiabatic Hamiltonian \cite{Demirplak2003Adiabatic,Demirplak2005Assisted,Demirplak2008Consistency,Berry2009Transitionless}. 
Thus, CD can be used to control, enhance, or suppress QTR between two subspaces. Consider a driven system Hamiltonian $\hat{H}_0(t)=\sum_nE_n(t)|n_t\rangle\langle n_t|$. When the system is prepared in an eigenstate $|n_0\rangle$, the adiabatic trajectory under slow driving reads $|\psi_n^{\rm ad}(t)\rangle =|n_t\rangle\exp[i\phi_n(t)]$, where $\phi_n(t)=-\int_0^tdt'E_n(t')/\hbar+i\int_0^tdt'\langle n_{t'}|d_{t'}n_{t'}\rangle$. CD guides the evolution along the adiabatic trajectory $|\psi_n^{\rm ad}(t)\rangle$ without relying on slow driving. The corresponding time-evolution operator admits the representation $\hat{U}_{\rm CD}=\sum_n|\psi_n^{\rm ad}(t)\rangle\langle n_0|$, and is generated by the CD Hamiltonian $\hat{H}_{\rm CD}(t)=\hat{H}_0(t)+\hat{H}_1(t)$ in the Schr\"odinger picture, where 
\begin{eqnarray}
\hat{H}_{1}(t)=i\hbar\sum_n\left[|d_tn_t\rangle\langle n_t|-\langle n_t|d_tn_t\rangle|n_t\rangle\langle n_t|\right].\nonumber
\end{eqnarray}
Under CD, the conditional probability and the QTR read
\begin{eqnarray}
P(B,t|A)&\!=\!&\sum_{n,m}\langle m_0|\hat{\rho}_A|n_0\rangle\langle n_t|\hat{\Pi}_B|m_t\rangle^{i(\phi_m(t)-\phi_n(t))},\qquad\\
k_{A\rightarrow B}(t)&\!=\!&\sum_{n,m}\langle m_0|\hat{\rho}_A|n_0\rangle \langle n_0|\hat{J}_B(t)|m_0\rangle,\quad
\end{eqnarray}
which is consistent with the identity (\ref{defkAB}). As shown in \cite{SM},
\begin{equation}
|k_{A\rightarrow B}(t)|\leq \frac{1}{\hbar}\sum_{n,m}\lvert\langle n_0|\hat{\rho}_A|m_0\rangle\rvert  \Delta_{|m_t\rangle}\hat H_{\rm CD}\Delta_{|n_t\rangle} \hat\Pi_B.
\end{equation}
The energy dispersion of the CD Hamiltonian $\hat{H}_{\rm CD}$ in a single eigenstate of $\hat{H}_0$ greatly simplifies \cite{delcampo12,Funo17}. In particular, consider the case of parametric driving, where $\hat H_0(t)=\hat H(\lambda_t)$, the vector $\lambda_t=(\lambda^1,\lambda^2,\cdots)$ specifies the time dependence of the Hamiltonian parameters and $|n_t\rangle=|n(\lambda_t)\rangle$. Then, 
$\Delta^2_{|n_t\rangle}\hat H_{\rm CD}(t)=\langle n_t|\hat{H}_1^2|n_t\rangle$. In addition, the last term can be written in terms of the real part of the quantum geometric tensor defined as \cite{Provost1980}, 
$Q_{\mu\nu}^{(n)}=\langle \partial_{\mu}n|(1-\hat{P}_n)|\partial_{\nu}n\rangle$,
associated with the $n$-th eigenstate of $\hat{H}_0$.
In terms of $g_{\mu\nu}^{(n)}(t)={\rm Re}[Q_{\mu\nu}^{(n)}]$, and using the semi-norm inequality \cite{Boixo07} to upper bound the variance of the projector as $\Delta^2_{|n_t\rangle} \hat\Pi_B(t)\leq 1/4$, it follows that the QTR satisfies the quantum geometric bound
\begin{equation}
|k_{A\rightarrow B}(t)| \leq \frac{1}{2}\sum_{n,m}\lvert\langle n_0|\hat{\rho}_A|m_0\rangle\rvert \sqrt{g_{\mu\nu}^{(m)}(t)\dot{\lambda}^\mu\dot{\lambda}^\nu},
\end{equation}
which further simplifies to a single sum when $\left[\hat{\rho}_A, \hat{H}_0(0)\right] = 0$.
Under CD, the QTR is thus upper bounded by an intrinsic geometric quantity, the weighted average of the line elements associated with each eigenstate $|n_t\rangle$, regardless of the definition of $\hat{\Pi}_B$. Further details and extensions to the local CD expansion for the TFIM are presented in the End Matter and in~\cite{SM}.

{\it Summary.} We have introduced a theoretical framework for the description of QTRs in arbitrary quantum processes. QTRs characterize the rate at which a quantum state moves from an initial subspace to a target subspace. As such,  they provide an alternative to QSLs in identifying the characteristic timescales of physical processes. QTRs can also be understood as a generalization of chemical reaction rates, applicable to arbitrary subspaces and dynamical processes, without resorting to a scattering approach and focusing instead on finite-time evolution. 
They can be formulated beyond Hamiltonian dynamics, encompassing both Markovian and non-Markovian processes, and incorporating the effects of quantum measurements as described by a quantum channel.
As such, we expect QTRs to find broad applications in nonequilibrium physics, ranging from the foundations of physics to quantum technologies. 

{\it Acknowledgements.} It is a pleasure to acknowledge discussions with \'I\~nigo L. Egusquiza and Norman Margolus.  
This project was supported by the Luxembourg National Research Fund (FNR Grant Nos.\ 17132054 and 16434093). It has also received funding from the QuantERA II Joint Programme and co-funding from the European Union’s Horizon 2020 research and innovation programme.  DEM was supported by the US National Science Foundation (Grant CHE 2400424) and the Alexander von Humboldt Foundation.
\bibliography{QRates}
\newpage

\section{End Matter}
{\it QSL by QTRs.}
In this section, we show that even in the simplest case of a single two-level system (TLS) with on-site energies $\pm\Delta$ and coupling term $W$, the QTR provides systematically tighter QSLs than the MT bound,
\begin{eqnarray}
    H=\Delta\left(\lvert 1\rangle\langle 1\rvert-\lvert 0\rangle\langle 0\rvert\right)+W\left(\lvert 0\rangle\langle 1\rvert+\lvert 1\rangle\langle 0\rvert\right)
\end{eqnarray}
with the system initialized in $\lvert 0\rangle$, $\lvert\psi(t)\rangle=c_-\lvert 0\rangle+c_+\lvert 1\rangle,\,c_-(0)=1,\,c_+(0)=0$. As shown in~\cite{SM}, the solution to the time-evolved state is given by 
\begin{eqnarray}
    c_-&=&\cos\left(\Omega t\right)+i\frac{\Delta}{\Omega}\sin\left(\Omega t\right),\\
    c_+&=&-i\frac{W}{\Omega}\sin\left(\Omega t\right),
\end{eqnarray}
with $\Omega=\sqrt{\Delta^2+W^2}$.
From here, the survival and transition probabilities become
\begin{eqnarray}
    &&P(A,t\vert A)=\cos^2(\Omega t)+\left(\frac{\Delta}{\Omega}\right)^2\sin^2(\Omega t),\\
    &&P(B,t\vert A)=\left(\frac{W}{\Omega}\right)^2\sin^2(\Omega t),
\end{eqnarray}
defining the MT and QTR-based QSLs, respectively,
\begin{eqnarray}
    \tau_\mathrm{MT}&\!=\!&\frac{\hbar \,\cos^{-1}\sqrt{P(A,\tau|A)}}{\frac{1}{\tau}\int_0^\tau \Delta_{\hat{\rho}_A} \hat H(t) dt}\\
    \tau_\mathrm{QTR}&\!=\!&\frac{\hbar \,\cos^{-1}\sqrt{P(B,\tau|A)d^{-1}_B+M_{\hat\rho_A(\tau)}\sqrt{1-d^{-1}_B}}}{\frac{1}{\tau}\int_0^\tau \Delta_{\hat{\rho}_A} \hat H(t) dt}.\quad
\end{eqnarray}
As shown in the main text, Eq.~\eqref{eq: QSLQTR}, the condition $P(B,t\vert A)\leq P(A,t\vert A)$ then captures the time domains with a tighter QSL by QTR,
\begin{eqnarray}
    &&|\Delta|\geq|W|,\,\,\tau_\mathrm{QTR}\geq\tau_{\mathrm{MT}},\,\,\forall t\in\mathbb R,\\
    &&\Delta\neq 0,\,\,\tau_\mathrm{QTR}\geq\tau_\mathrm{MT},\,\, t\in[0,t^*]+\mathrm{mod}(\pi/(2\Omega)),\\
    &&t^*\geq \pi/(4\Omega).\nonumber
\end{eqnarray}
Here, the first line shows clearly that whenever the on-site energy difference exceeds that of the couplings, the QTR-based QSL is tighter for all $t$, while the second line implies that for a larger fraction in each period, the QTR-based QSL is tighter for any non-zero $\Delta$.

\begin{figure}
    \centering
    \includegraphics[width=.9\columnwidth]{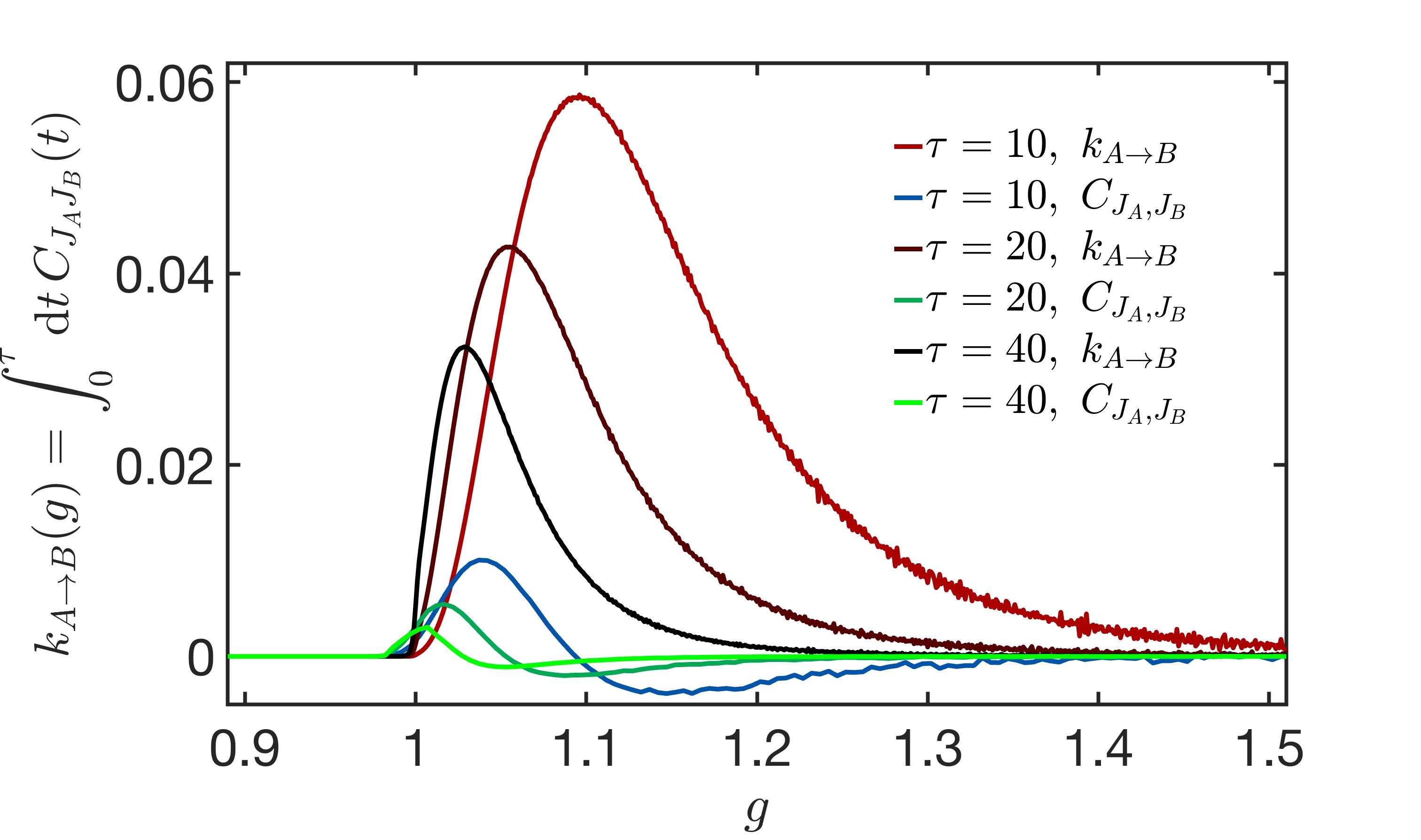}
    \caption{QTRs and flux-flux correlation functions for the TFIM for various driving times exhibiting a sharp change around the critical point.}\label{fig: kAB_CAB_TFIM} 
\end{figure}

{\it Flux-flux correlation approach in the TFIM.} In this section, we show how the study of QTRs in the TFIM can be facilitated by the use of the flux-flux correlation function. The corresponding Hamiltonian is given by~\cite{Zurek2005Dynamics,Dziarmaga2005Dynamics,Suzuki2012Quantum,Sachdev2011Quantum}
\begin{eqnarray}
    \hat{H}(t)\!&=&\!-\!\sum_{j=1}^L \left( \hat{\sigma}^z_j \hat{\sigma}^z_{j+1} + g(t) \hat{\sigma}_j^x \right),\\g(t)\!&=&\!g(0)(1-t/\tau),\quad\quad
\end{eqnarray}
with $\hat\sigma^{z,x}_j$ denoting the Pauli spin-$1/2$ operators at the $j$-th site. 
Using the Jordan-Wigner transformation of the spin operators, $\hat\sigma^x_j=2\hat c^\dagger_j\hat c_j-1$ and $\hat\sigma^z_j=\prod_{j=1}^{k-1}\hat\sigma^x_j(\hat c^\dagger_j+\hat c_j)/2$ with $\hat c^\dagger_j,\,\hat c_j$ being fermionic creation and annihilation operators, and a subsequent Fourier decomposition of $\hat c_k=e^{-i\pi /4}\sum_j e^{-i\pi jk}\hat c_j$, leads to the independent TLS representation,
\begin{eqnarray}
    \hat{H} 
    = 2 \sum_{k>0} \hat{\psi}_k^\dagger \left[ (g(t)-\cos k)\hat \sigma^z_k + \sin k \, \hat \sigma^x_k \right] \hat{\psi}_k,
\end{eqnarray}
with $\hat \sigma^{x,y,z}_k$ denoting the Pauli matrices of the $k$-th TLS.
Here, $\hat{\psi}_k := (\hat{c}_k, \hat{c}_{-k}^\dagger)^T$ is a vector of creation and annihilation operators for fermions of momentum $k$, and $\tau^{x,y,z}$ are another set of Pauli matrices, acting on the internal space of $\hat{\psi}_k$. Consider the target subspace associated with the low-energy subspace of all modes up to a given cut-off momentum $k_B$ and the initial ground state for $\hat\rho_A$. The flux-flux correlation function is given by~\cite{Zurek2005Dynamics,Dziarmaga2005Dynamics,Suzuki2012Quantum,Sachdev2011Quantum}
\begin{eqnarray}
    &&C_{AB}(t)=\tr[\hat\rho_A\hat J_B(t)\hat J_{A,t}],\\
    &&\hat J_{B}(t)=(i\hbar)^{-1}\left[\hat\Pi_B(t),\hat H(t)\right],\,\hat J_{A,t}=(i\hbar)^{-1}\left[\hat\Pi_A,\hat H(t)\right],\nonumber
\end{eqnarray}
with $\hat\rho_A=\prod_{k>0}\lvert 0\rangle_{k}\lvert 0\rangle_{-k}\langle 0\rvert_{-k}\langle 0\rvert_{k}$ and
\begin{eqnarray}\label{eq: rho_A_Pi_B_t}    
    &&\hat\Pi_B=\prod_{k_B>k>0}\big[\sin^2\frac{k}{2}\lvert0\rangle_k\lvert0\rangle_{-k}\langle0\rvert_{-k}\langle0\rvert_k\nonumber\\
    &&-\frac{\sin k}{2}\Big(\lvert1\rangle_k\lvert1\rangle_{-k}\langle0\rvert_{-k}\langle0\rvert_k+\lvert0\rangle_k\lvert0\rangle_{-k}\langle1\rvert_{-k}\langle1\rvert_k\Big)\nonumber\\
    &&+\cos^2\frac{k}{2}\lvert1\rangle_k\lvert1\rangle_{-k}\langle1\rvert_{-k}\langle1\rvert_k\Big],
\end{eqnarray}
where $\lvert 0\rangle_k,\,\lvert 1\rangle_k$ are the eigenstates of the particle number operator $\hat c^\dagger_k\hat c_k$ of the $k$-th mode.
 As detailed in~\cite{SM}, this correlation function becomes
\begin{eqnarray}
    &&C_{AB}(t)=\sum_{k>0}h_{12,k}\left[2h_{11,k}u_{12,k}-h_{12,k}\left(u_{11,k}-u_{22,k}\right)\right]\!,\qquad\\
    &&u_{11,k}\!=\!\lvert v_k\rvert^2\sin^2\frac{k}{2}-(u_kv^*_k+v_ku^*_k)\sin  k+|u_k|^2\cos^2\frac{k}{2},\nonumber\\
    &&u_{12,k}=\frac{1}{2}\sin k\left((|u_k|^2-|v_k|^2\right)-\cos^2\frac{k}{2}v_ku_k+u^*_kv^*_k\sin^2\frac{k}{2},\nonumber\\
    &&u_{22,k}\!=\!|u_k|^2\sin^2\frac{k}{2}+\sin  k(u^*_kv_k+v^*_ku_k)+|v_k|^2\cos^2\frac{k}{2},\nonumber
\end{eqnarray}
with $h_{11,k},\,h_{12,k}$ are further  complicated fucntions of $u_k$ and $v_k$ (see~\cite{SM}).
Here, the $u_k$ and $v_k$ coefficients are defined as the solutions of the time-dependent Schr\"odinger equation of the $k$-th mode, $\hat U_k(t)\lvert0\rangle_k\lvert0\rangle_{-k}=u_k\lvert1\rangle_k\lvert1\rangle_{-k}+v_k\lvert0\rangle_k\lvert0\rangle_{-k}$.
Knowledge of the closed-form expression for the flux-flux correlator can thus be used for the efficient computation of the QTR in the TFIM rather than by differentiating the transition probability. By way of example, Fig.~\ref{fig: kAB_CAB_TFIM}
reports the QTRs computed via the flux-flux correlator in the TFIM for a linear quench with different driving times. The critical dynamics is reflected in the sharp growth of the  QTR upon crossing the critical point $g_c=1$. As the driving rate $1/\tau$ increases, the flux-flux correlation function and QTR vary more rapidly.

{\it Counterdiabatic Driving.}
In this section, we detail the main results on the QTRs in the presence of the Krylov expansion of the CD in the TFIM, which has recently been implemented experimentally~\cite {Visuri_2025}.
We employ the Krylov expansion of the exact CD in momentum space,
\begin{eqnarray}
    \hat H^{(n)}_1(t)=\sum_{k>0}q^{(n)}_k(t) \hat{\psi}^\dagger_k\hat\sigma^y_k\hat{\psi}_k,
\end{eqnarray}
where $q^{(n)}_k(t)=-\frac{1}{\tau}\sum_{m=1}^n\frac{\sin(km)}{2}\frac{g^{2m}(1+g^L)}{g^{m+1}+g^L}$. Starting the system from the initial ground state at $g$ with the target subspace given by the low energy subspace up to momentum $k_B$,
$\hat\rho_A(\tau)=\prod_{k>0} \left(u_k(\tau)\lvert 0\rangle_{k}\lvert 0\rangle_{-k}+v_k(\tau)\lvert 1\rangle_{k}\lvert 1\rangle_{-k}\right)
\times\left(u^*_k(\tau)\langle 0\rvert_{-k}\langle 0\rvert_{k}+v^*_k(\tau)\langle 1\rvert_{-k}\langle 1\rvert_{k}\right)$,
with $\hat\Pi_B$ given by Eq.~\eqref{eq: rho_A_Pi_B_t}.
In the sudden quench limit, $\tau\rightarrow 0$, the CD-assisted time evolution can be solved exactly with the coefficient given by
\begin{eqnarray}
    &&u^{(n)}_k=\sin\frac{k}{2}\cos\left[\sum_{m=0}^\infty\frac{\sin[k(n+m+1)]}{n+m}\right]\nonumber\\
    &&-\cos\frac{k}{2}\sin\left[\sum_{m=0}^\infty\frac{\sin[k(n+m+1)]}{n+m}\right],\\
    &&v^{(n)}_k=-\cos\frac{k}{2}\cos\left[\sum_{m=0}^\infty\frac{\sin[k(n+m+1)]}{n+m}\right]\nonumber\\
    &&+\sin\frac{k}{2}\sin\left[\sum_{m=0}^\infty\frac{\sin[k(n+m+1)]}{n+m}\right],
\end{eqnarray}
with the initial condition $v^{(n)}_k=0,\,u^{(n)}_k=1$.
For large $n$ one finds~\cite{grabarits2025UniversalCD},
    $v^{(n)}_k\approx-\sin\left[\mathrm{Si}(nk)-k/2\right],\quad u^{(n)}_k\approx\cos\left[\mathrm{Si}(nk)-k/2\right],\,$
with $\mathrm{Si}(x)=\int_0^x\mathrm dz\,\frac{\sin(z)}{z}$ is the sine-integral function.
The transition probability in terms of $u^{(n)}_k$ and $v^{(n)}_k$ can be written as
\begin{eqnarray}\label{eq: rho_A(t)}
    &&P(B,\tau\vert A)\!=\!\tr[\hat\rho_A(\tau)\hat\Pi_B]=\prod_{0<k<k_B} \left\lvert v_k\sin \frac{k}{2}-u_k\cos \frac{k}{2} \right\rvert^2,\nonumber\\
\end{eqnarray}
where the trace in the transition probability was restricted to the product of the excited states up to momentum $k\leq k_B$ by $\hat\Pi_B$.
\begin{figure}[t]
\centering    \includegraphics[width=.9\columnwidth,trim={0 5cm 0 0},clip]{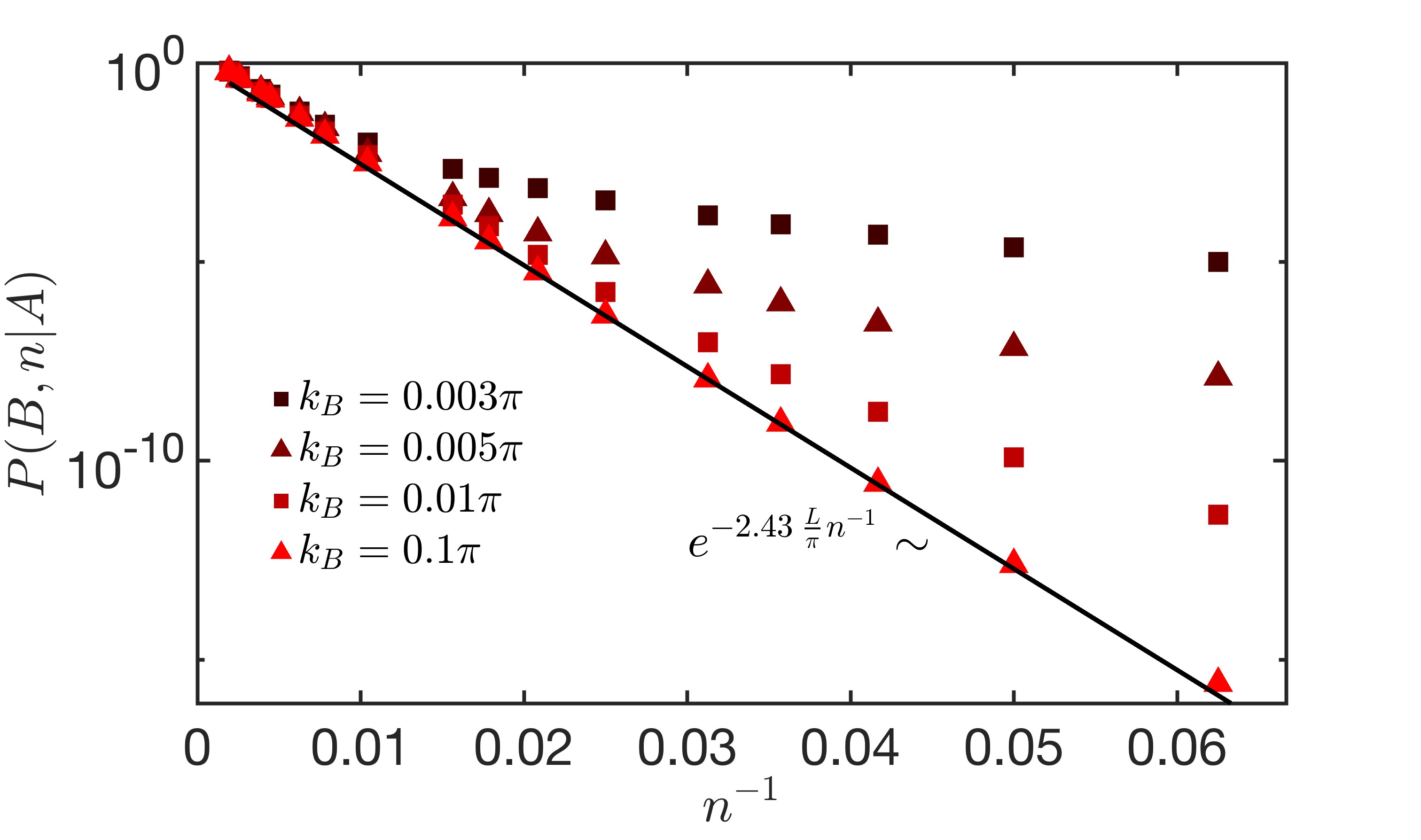}
    \includegraphics[width=.9\columnwidth]{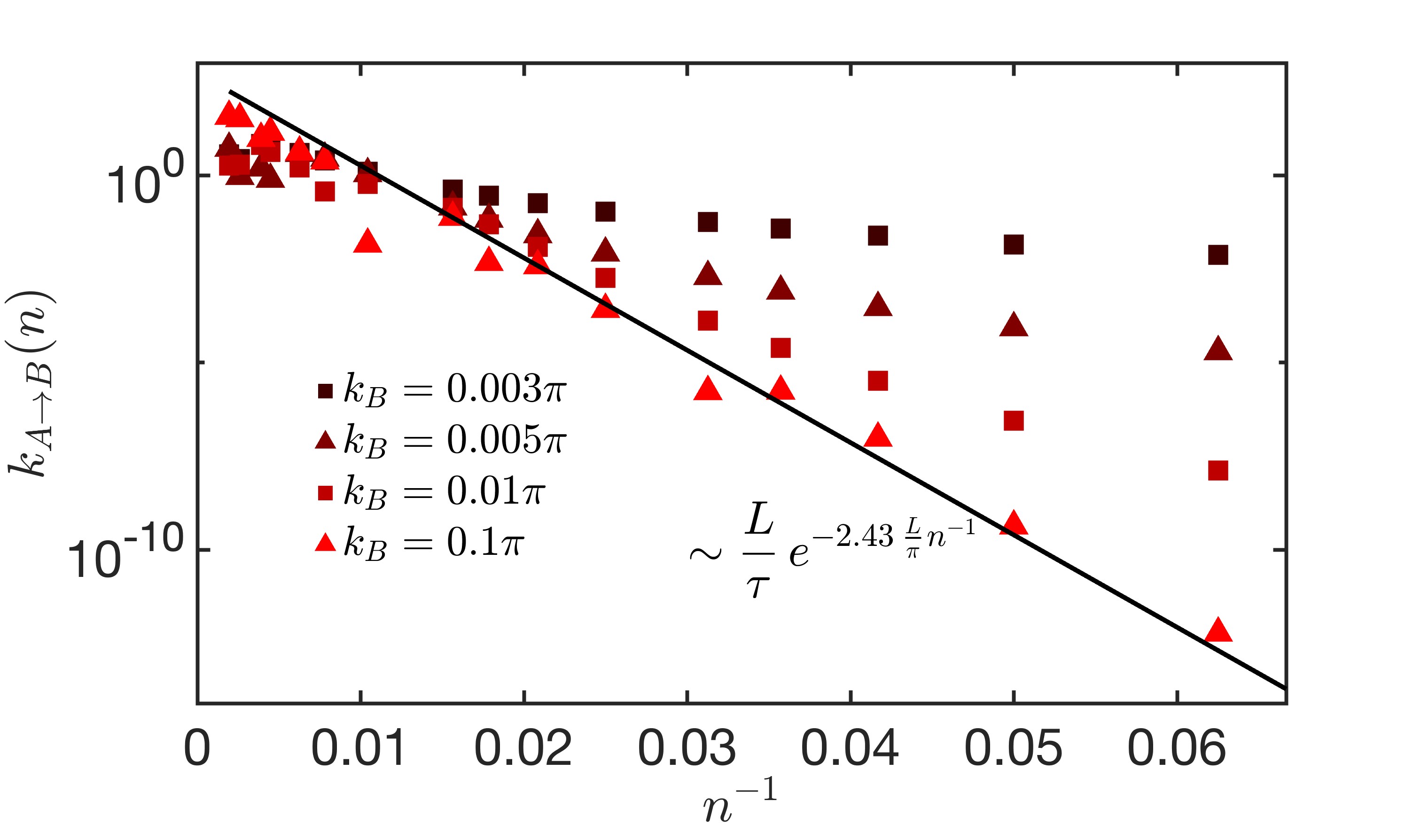}
    \caption{(Up) Transition probability in the TFIM under CD, precisely following the theoretical prediction for a broad range of cut-off momenta. ($L=1600$). (Down) QTR with similar agreement wit the analytical predictions.}\label{fig:PAB_CD}
\end{figure}
Using the results of Ref.~\cite{grabarits2025UniversalCD}
$\left\lvert u^{(n)}_k\sin \frac{k}{2}-v^{(n)}_k\cos \frac{k}{2} \right\rvert^2\approx \sin^2\left[\mathrm{Si}(nk)\right]$, in the sudden limit,
\begin{eqnarray}\label{eq: P_AB_CD}
    P(B,n\vert A)&\approx&\prod_{k<k_B}\sin^2\left[\mathrm{Si}(nk)\right]
    \approx e^{\frac{2L}{\pi}\int_0^{k_B}\mathrm dk\,\log\left[\lvert\sin(\mathrm{Si}(nk))\rvert\right]}\nonumber\\
    &=&e^{\frac{2L}{\pi n}\int_0^{\infty}\mathrm dk\,\log\left[\lvert\sin(\mathrm{Si}(x))\rvert\right]}\approx e^{-2.48\frac{L}{\pi}n^{-1}},
\end{eqnarray}
provided that $n\gg 1/k_B$, and where we have switched to the more convenient notation, $P(B,n\vert A)$. Its time-derivative yields the transition rate,
\begin{eqnarray}
    k_{A\rightarrow B}(n)\!&=&\frac{\mathrm d P(B,n\vert A)}{\mathrm dt}\\
    &=&2\mathrm{Re}\sum_{k<k_B}\frac{ \left(\dot v^{(n)}\right)^*_{k^\prime}\sin\frac{k'}{2}-\left(\dot u^{(n)}\right)^*_{k^\prime}\cos\frac{k'}{2}}{\left(v^{(n)}\right)^*_{k^\prime}\sin\frac{k'}{2}-\left(u^{(n)}\right)^*_{k^\prime}\cos\frac{k'}{2}}\nonumber\\
    &&\times\prod_{k'<k_B}\left\lvert v^{(n)}_k\sin \frac{k'}{2}-u^{(n)}_k\cos \frac{k'}{2} \right\rvert^2\sim \frac{L}{\tau}e^{-2.48\frac{L}{\pi}n^{-1}},\nonumber
\end{eqnarray}
As shown in Fig.~\ref{fig:PAB_CD}, the domain of perfect agreement with numerical results extends to larger target subspaces in momentum space.

\newpage \appendix
\clearpage
\widetext
%\minitoc

\begin{center}
{\bf Supplementary Information for:\\``Quantum Transition Rates in Arbitrary Physical Processes''}
\end{center}
\setcounter{tocdepth}{1}   % 1 = sections only; 2 = include subsections
\tableofcontents
\clearpage
%\localtableofcontents 
%\cleardoublepage
%\tableofcontents
%\cleardoublepage
\section{QTR from generalized flux-flux correlator}

Let us evaluate the time derivative of the QTR explicitly
\begin{eqnarray}
\frac{d}{dt}k_{A\rightarrow B}(t)
&=&
\tr[\hat{\rho}_A\frac{d}{dt}\hat{J}_B(t)]
\nonumber\\
&=&
\tr\left[
\hat{\rho}_A \left(
\frac{1}{i\hbar}[\hat{J}_B(t),\hat{H}(t)]
+
\frac{1}{i\hbar}[\hat{\Pi}_B \left( t \right), \dot{\hat{H}} \left( t \right)] 
\right) 
\right]
.
\end{eqnarray}
The first term can alternatively be rewritten as
\begin{eqnarray}
\frac{1}{i\hbar}
\tr\left(
\hat{\rho}_A 
[\hat{J}_B(t),\hat{H}(t)]
\right)
&=&
\frac{1}{i\hbar}\tr\left(\hat{\Pi}_A\hat{\rho}_A\hat{\Pi}_A[\hat{J}_B(t),\hat{H}(t)]\right)\nonumber\\
&=&-\frac{1}{i\hbar}\tr\left([\hat{\Pi}_A\hat{\rho}_A\hat{\Pi}_A,\hat{H}(t)]\hat{J}_B(t)\right)\nonumber\\
&=&-\frac{1}{i\hbar}\tr\left([\hat{\Pi}_A,\hat{H}(t)]\hat{\rho}_A\hat{\Pi}_A\hat{J}_B(t)\right)
-\frac{1}{i\hbar}\tr\left(\hat{\Pi}_A[\hat{\rho}_A,\hat{H}(t)]\hat{\Pi}_A\hat{J}_B(t)\right)\nonumber\\
& &
-\frac{1}{i\hbar}\tr\left(\hat{\Pi}_A\hat{\rho}_A[\hat{\Pi}_A,\hat{H}(t)]\hat{J}_B(t)\right)
\nonumber\\
&=&-\tr\left(
%\hat{J}_A
\hat{J}_{A, t} 
\hat{\rho}_A\hat{J}_B(t)\right)
-\frac{1}{i\hbar}\tr\left(\hat{\Pi}_A[\hat{\rho}_A,\hat{H}(t)]\hat{\Pi}_A\hat{J}_B(t)\right) -
\tr\left(\hat{\rho}_A
%\hat{J}_A
\hat{J}_{A, t} 
\hat{J}_B(t)\right)
\nonumber\\
&=&
-2\tr\left(\hat{J}_B(t)
%\hat{J}_A
\hat{J}_{A, t} 
\hat{\rho}_A\right)-\frac{1}{i\hbar}\tr\left(\hat{\Pi}_A[\hat{\rho}_A,\hat{H}(t)]\hat{\Pi}_A\hat{J}_B(t)\right)
+ 
\tr \left( 
\hat{J}_B \left( t \right) 
\left\lbrack 
%\hat{J}_A 
\hat{J}_{A, t} 
, 
\hat{\rho}_A 
\right\rbrack 
\right) 
.
\end{eqnarray}

Since 
$
%\hat{J}_A 
\hat{J}_{A, t} 
\coloneqq 
\left\lbrack 
\hat{\Pi}_A 
, 
\hat{H} \left( t \right) 
\right\rbrack / (i \hbar)
$, it follows that
\begin{eqnarray}
i \hbar 
\left\lbrack 
%\hat{J}_A 
\hat{J}_{A, t} 
, 
\hat{\rho}_A 
\right\rbrack 
& = & 
\hat{\Pi}_A 
\hat{H} \left( t \right) 
\hat{\rho}_A 
- 
\hat{H} \left( t \right) 
\hat{\Pi}_A 
\hat{\rho}_A 
- 
\hat{\rho}_A 
\hat{\Pi}_A 
\hat{H} \left( t \right) 
+ 
\hat{\rho}_A 
\hat{H} \left( t \right) 
\hat{\Pi}_A 
\nonumber\\
& = & 
\hat{\Pi}_A 
\hat{H} \left( t \right) 
\hat{\rho}_A 
- 
\hat{H} \left( t \right) 
\hat{\rho}_A 
\hat{\Pi}_A 
- 
\hat{\Pi}_A 
\hat{\rho}_A 
\hat{H} \left( t \right) 
+ 
\hat{\rho}_A 
\hat{H} \left( t \right) 
\hat{\Pi}_A 
\nonumber\\
& = & 
- 
\left\lbrack 
\left\lbrack 
\hat{H} \left( t \right) 
, 
\hat{\rho}_A 
\right\rbrack 
, 
\hat{\Pi}_A 
\right\rbrack 
, 
\end{eqnarray}
giving 
\begin{eqnarray}\label{eq: k_transformation}
\frac{d}{d t} 
k_{A \rightarrow B} \left( t \right) 
&=& 
- 
2 \tr\left(\hat{J}_B(t)
%\hat{J}_A
\hat{J}_{A, t} 
\hat{\rho}_A\right)
+
\frac{1}{i\hbar}
\tr\left(
\hat{J}_B(t) \hat{\Pi}_A[\hat{H}(t), \hat{\rho}_A]\hat{\Pi}_A
\right)
\nonumber\\
& &
- 
\frac{1}{i \hbar} 
\tr \left( 
\hat{J}_B \left( t \right) 
\left\lbrack 
\left\lbrack 
\hat{H} \left( t \right) 
, 
\hat{\rho}_A 
\right\rbrack 
, 
\hat{\Pi}_A 
\right\rbrack 
\right) 
+ 
\frac{1}{i \hbar} 
\tr \left( 
\hat\rho_A
\left\lbrack 
\hat{\Pi}_B \left( t \right) 
, 
\dot{\hat{H}} \left( t \right) 
\right\rbrack 
\right) 
. 
\end{eqnarray}
Furthermore, assuming that $[\hat\rho_A,\hat H_S(t)]=0$ and $\left[\hat\rho_A,\hat U(t)\right]=0$, the last term
in Eq.~\eqref{eq: k_transformation} vanishes by the relation
\begin{eqnarray}
    \tr \left(\hat\rho_A
\left\lbrack 
\hat{\Pi}_B \left( t \right) 
, 
\dot{\hat{H}} \left( t \right) 
\right\rbrack 
\right)=\tr \left(\hat\rho_A
\left\lbrack 
\hat{\Pi}_B
, 
\dot{\hat{H}}_S \left( t \right) 
\right\rbrack 
\right)=\frac{d}{dt}\tr \left(\hat\rho_A
\left\lbrack 
\hat{\Pi}_B
, 
\hat{H}_S \left( t \right) 
\right\rbrack 
\right)=-\frac{d}{dt}\tr \left(
\left\lbrack 
\hat\rho_A
, 
\hat{H}_S \left( t \right) 
\right\rbrack \hat{\Pi}_B
\right)=0.
\end{eqnarray}
Integrating term by term, leads to the expression of the QTR as given in the main text, Eq. \eqref{eq:k_fluxflux}%(11)
. The simplified relation in Eq. \eqref{fluxfluxeq} %(12)
follows from the assumptions $
\left\lbrack 
\hat{\rho}_A, 
\hat{H}_S(t)
\right\rbrack 
= 
0 
$ and 
$
\hat{\Pi}_A = \mathbb{I} - \hat{\Pi}_B
$. 
We note that the flux-flux correlation vanishes at $t=0$ as 
$
C_{J_BJ_B} \left( 0 \right) 
= 
\tr \left\lbrack 
\hat{J}_B^2 \left( 0 \right) 
\hat{\rho}_A 
\right\rbrack 
= 
\tr \left\lbrack 
\hat{H} \left( 0 \right) 
\hat{\Pi}_B 
\hat{\rho}_A 
\hat{H} \left( 0 \right) 
\right\rbrack / \hbar^2 
= 
0 
$.

The time-independent case naturally follows from the above results by setting the Hamiltonian to $\hat H(t)\rightarrow\hat H$ and so dropping the explicit time-derivative part, $\dot{\hat H}=0$. Furthermore, the generalized flux operator also simplifies to the standard flux operator at $t=0$, $\hat J_{A,t}\rightarrow \hat J_A\coloneqq \hat J_A(0)=\left[\hat\Pi_A,\hat H\right]/(i\hbar)$. As a result, the QTR can be written as
\begin{eqnarray}%\label{eq: k_transformation}
\frac{d}{d t} 
k_{A \rightarrow B} \left( t \right) 
&=& 
- 
2 \tr\left(\hat{J}_B(t)
%\hat{J}_A
\hat{J}_A
\hat{\rho}_A\right)
+
\frac{1}{i\hbar}
\tr\left(
\hat{J}_B(t) \hat{\Pi}_A[\hat{H}, \hat{\rho}_A]\hat{\Pi}_A
\right)
- 
\frac{1}{i \hbar} 
\tr \left( 
\hat{J}_B \left( t \right) 
\left\lbrack 
\left\lbrack 
\hat{H} 
, 
\hat{\rho}_A 
\right\rbrack 
, 
\hat{\Pi}_A 
\right\rbrack 
\right).  
\end{eqnarray}
Integrating all terms, whenever $\left[\hat\rho_A,\hat H\right]=0$ and $\hat\Pi_A=\mathbb I-\hat\Pi_B$, leads to the generalized flux-flux correlation form, Eq.~\eqref{fluxfluxeq_timeindep} given that $C_{J_BJ_B}(0)=0$.

\section{Two-level system example for tighter QTR-based QSL}
We consider a simple two level system with energies $-\Delta$ and $\Delta$ and with coupling term of $W$, given by the Hamiltonian,
\begin{eqnarray}
    H=\Delta\left(\lvert 1\rangle\langle 1\rvert-\lvert 0\rangle\langle 0\rvert\right)+W\left(\lvert 0\rangle\langle 1\rvert+\lvert 1\rangle\langle 0\rvert\right),
\end{eqnarray}
with the system initialized in $\lvert 0\rangle$ with $c_-(0)=1,\,c_+(0)=0$. The corresponding set of differential equations that describes their evolution is given by
\begin{eqnarray}
    \dot c_-&=&i\Delta c_--iW c_+,\\
    \dot c_+&=&-iWc_--i\Delta c_+,
\end{eqnarray}
which is readily solved by
\begin{eqnarray}
    c_-&=&\cos\left(\Omega t\right)+i\frac{\Delta}{\Omega}\sin\left(\Omega t\right),\\
    c_+&=&-i\frac{W}{\Omega}\sin\left(\Omega t\right).
\end{eqnarray}
The corresponding survival and transition probabilities for the time-evolved state $\lvert\psi(t)\rangle=c_-\lvert0\rangle+c_+\lvert1\rangle$, are given by
\begin{eqnarray}
    &&P(A,t\vert A)=\lvert\langle\psi(0)\vert\psi(t)\rangle\rvert^2=\lvert c_-\rvert^2=\cos^2(\Omega t)+\left(\frac{\Delta}{\Omega}\right)^2\sin^2(\Omega t),\\
    &&P(B,t\vert A)=\lvert\langle\psi(t)\vert1\rangle\rvert^2=\lvert c_+\rvert^2\equiv 1-\lvert c_-\rvert^2=\left(\frac{W}{\Omega}\right)^2\sin^2(\Omega t),
\end{eqnarray}
which define the MT and QTR-based QSLs as
\begin{eqnarray}
    \tau_\mathrm{MT}&\!=\!&\frac{\hbar \,\cos^{-1}\sqrt{P(A,\tau|A)}}{\frac{1}{\tau}\int_0^\tau \Delta_{\hat{\rho}_A} \hat H(t) dt},\\
    \tau_\mathrm{QTR}&\!=\!&\frac{\hbar \,\cos^{-1}\sqrt{P(B,\tau|A)d^{-1}_B+M_{\hat\rho_A(\tau)}\sqrt{1-d^{-1}_B}}}{\frac{1}{\tau}\int_0^\tau \Delta_{\hat{\rho}_A} \hat H(t) dt}.\quad
\end{eqnarray}
The tightness competition between the MT and QTRs-based QSLs is captured by the relation $P(B,t\vert A)\leq P(A,t\vert A)$,
\begin{eqnarray}
    &&P(B,t\vert A)\leq P(A,t\vert A)\Rightarrow 1- \lvert c_-\rvert^2\leq \lvert c_-\rvert^2\\
    &&\Rightarrow \lvert c_-\rvert^2=\frac{\Delta^2+W^2\cos^2(\Omega t)}{\Delta^2+W^2}\geq 1/2\Rightarrow \cos^2\left(\Omega t\right)\geq\frac{1}{2}\left(1-\frac{\Delta^2}{W^2}\right),\nonumber
\end{eqnarray}
which is satisfied in the simplest way if $\Delta>W$, i.e., for on-site energy differences larger than the ones of the couplings. A more refined bound can be obtained by realizing that for every positive $\Delta\neq 0$, it holds that $\cos^2(t\,\Omega)\geq1/2$ for every $t\Omega\in[0,\pi/4]+\mathrm{mod}(\pi)$, i.e., for the majority of the elapsed time.

\section{Example:  QTRs in the Bixon-Jortner model}
By way of example, we consider the QTRs in the celebrated Bixon-Jortner model, describing the intramolecular radiationless transitions in an isolated molecule. This will help illustrate that the formulation of QTRs is not restricted to continuous variables but is also applicable to quantum systems with discrete degrees of freedom. In addition, we shall discuss the QTRs as a function of the choice of the target subspace. 

In this model, an isolated energy level, $\lvert 0\rangle$ with zero energy and occupation amplitude denoted by $b(t)$ is coupled to infinitely many states with uniformly spaced energy levels, $\lvert n\rangle,\,E_n=n\Delta $. Thus, $\Delta$ sets the level spacing. The occupation amplitude in each of the infinitely many states is denoted by $c_n$.  When the system is initialized in the isolated level, $b(0)=1,\,c_n(0)=0$, the time evolution of the occupation amplitudes can be written as
\begin{eqnarray}
    &&H=\sum_nE_n\lvert n\rangle\langle n\rvert+\sum_n W_n\left(\lvert n\rangle \langle 0\rvert+\lvert 0\rangle \langle n\rvert\right)\\
    &&\dot b(t)=-i\sum_n\,W_nc_n(t),\\
    &&\dot c_n(t)=-i n\Delta c_n(t)-i W_nb(t),
\end{eqnarray}
for general non-flat couplings, $W_n$.
To extract analytical properties, we consider the continuum limit with flat couplings $W_n=W$, where $c_n(t)\rightarrow c_f(t),\,\lvert n \rangle\rightarrow\lvert f\rangle,\,E_n\rightarrow\Delta_f$ is labeled by the continuous variable $f$ and $b(t)\rightarrow c_1(t)$. The time evolution can be expressed in terms of the Laplace transforms
\begin{eqnarray}
    &&s\overline{b}(s)-1=-iW\int\frac{\mathrm d\Delta_f}{\Delta}\overline c_{f}(s),\\
    &&s\overline c_{f}(s)=-i\Delta_{f}\overline c_{f}(s)-iW\overline b(s).
\end{eqnarray}
The solution is obtained again in terms of the Laplace transforms,
\begin{eqnarray}
    \overline {b}(s)=\frac{1}{s+\int\frac{\mathrm d\Delta_f}{\Delta}\frac{W^2}{s+i\Delta_f}}.
\end{eqnarray}
For an unbounded energy spectrum, one finds
\begin{eqnarray}
    \overline{b}(s)=\frac{1}{s+\frac{\pi W^2}{\Delta}}\Rightarrow b(t)=e^{-\frac{\pi W^2t}{\Delta}}.
\end{eqnarray}
In this limit, one can also express the transition amplitudes,
\begin{eqnarray}
    \overline c_f(s)=-iW\frac{1}{s+i\Delta_f}\frac{1}{s+\frac{\pi W^2}{\Delta}}.
\end{eqnarray}
The inverse Laplace transform by the Cauchy formula yields the following result
\begin{eqnarray}
    c_f(t)=\frac{1}{2\pi i}\int_{\gamma-i\,\infty}^{\gamma+i\,\infty}\mathrm ds\,\overline c_f(s)=-iW\frac{e^{-i\Delta_ft}-e^{-\frac{\pi W^2t}{\Delta}}}{\frac{\pi W^2}{\Delta}-i\Delta_f},
\end{eqnarray}
for an arbitrary positive real $\gamma$.

To this end, we consider QTRs with the initial density matrix given by the pure state of the isolated level and the target subspace parametrized by an arbitrary energy window $\delta E$,
\begin{eqnarray}
    &&\hat\rho_A=\lvert 0\rangle\langle 0\rvert,\\
    &&\hat\Pi_B=\frac{1}{\Delta}\int_{\Delta_f\in \delta E}\mathrm d\Delta_f\lvert f\rangle\langle f\rvert,
\end{eqnarray}
where the energy interval of the target subspace was parametrized as $[E_0,E_0+(E_1-E_0)\delta E],\,\delta E\in[0,1]$.

Using the expansion of the coefficients, the time-evolved density matrix is given by
\begin{eqnarray}
    \hat\rho_A(t)=\lvert 0(t)\rangle\langle 0(t)\rvert=\lvert b(t)\rvert^2\lvert 0\rangle\langle 0\rvert +\frac{1}{\Delta^2}\int\mathrm d\Delta_f\mathrm d\Delta_{f^\prime}\lvert f\rangle\langle f^\prime\rvert c_f(t)c^*_{f^\prime}(t)+\frac{1}{\Delta}\int\mathrm d\Delta_f c_f(t)\lvert f\rangle\langle 0\rvert +c^*_f(t)\lvert 0\rangle\langle f\rvert.
\end{eqnarray}
With this representation, in the transition probability, only the diagonal terms survive and
\begin{eqnarray}\label{eq: PAB}
    P(B,t\vert A)&=&\mathrm{Tr}(\hat\rho_A(t)\,\hat\Pi_B)=\frac{1}{\Delta}\int_{\Delta_f\in\delta E}\mathrm d\Delta_f\lvert c_f(t)\rvert^2=\frac{1}{\Delta}\int_{E_0}^{E_1}\mathrm d\Delta_f\left\lvert W\frac{e^{-i\Delta_ft}-e^{-\frac{\pi W^2t}{\Delta}}}{\frac{i\pi W^2}{\Delta}+\Delta_f}\right\rvert^2\\
    &=&\frac{1}{\Delta}\int_{E_0}^{E_1}\mathrm d\Delta_fW^2\frac{1-2e^{-\frac{\pi W^2t}{\Delta}}\cos(\Delta_ft)+e^{-\frac{2\pi W^2t}{\Delta}}}{\frac{\pi^2 W^4}{\Delta^2}+\Delta^2_f}\nonumber\\
    &=&\frac{1+e^{-\frac{2\pi W^2t}{\Delta}}}{\pi}\left[\arctan\left(\frac{E_1\Delta}{\pi W^2}\right)-\arctan\left(\frac{E_0\Delta}{\pi W^2}\right)\right]\nonumber\\
    &&-\frac{1}{\pi}\mathrm{Im}\Big[e^{-\frac{\pi W^2}{\Delta}t}\left[\mathrm{Ei}(itE_1+t\pi W^2/\Delta )-\mathrm{Ei}(itE_0+t\pi W^2/\Delta)\right]\nonumber\\
    &&-\left[\mathrm{Ei}(itE_1-t\pi W^2/\Delta)-\mathrm{Ei}(itE_0-t\pi W^2/\Delta)\right]\Big],\nonumber
\end{eqnarray}
where the exponential integral function has been introduced as $\mathrm{Ei}(z)=-\int_{-\infty}^z\mathrm dx\frac{e^{x}}{x}$. The first term comes from the standard integral,
\begin{eqnarray}
    \int_{E_0}^{E_1}\mathrm d\Delta_f\,\frac{1}{\Delta^2_f+\frac{\pi^2W^4}{\Delta^2}}=\arctan\left(\frac{\Delta E_1}{\pi W^2}\right)-\arctan\left(\frac{\Delta E_0}{\pi W^2}\right).
\end{eqnarray}
The exponential term with $\gamma=\pi W^2/\Delta$ is rewritten as the real part of 
\begin{eqnarray}
    \mathrm{Re}\int_{E_0}^{E_1}\mathrm d\Delta_f\frac{e^{i\Delta_f t}}{\Delta^2_f+\gamma^2}&=&\frac{1}{2\gamma}\mathrm{Im}\int_{E_0}^{E_1}\mathrm d\Delta_fe^{i\Delta_f t}\left(\frac{1}{\Delta_f-i\gamma}-\frac{1}{\Delta_f+i\gamma}\right)\\
    &=&\mathrm{Im}\left[\frac{e^{-\gamma t}}{\pi\gamma }\left[\mathrm{Ei}(itE_1+\gamma t)-\mathrm{Ei}(itE_0+\gamma t)\right]-\frac{1}{\pi\gamma }\left[\mathrm{Ei}(itE_1-\gamma t)-\mathrm{Ei}(itE_0-\gamma t)\right]\right]\nonumber
\end{eqnarray}
where we used the definition of $\mathrm{Ei}(x)=\int_{-\infty}^x\mathrm dz\frac{e^z}{z}$ and $\int_{E_0}^{E_1}\mathrm dx \frac{e^{ixt}}{x-i\gamma}=\int_{E_0-i\gamma}^{E_1-i\gamma}\mathrm dx \frac{e^{i(x+i\gamma)t}}{x}=e^{-\gamma t}\left[\mathrm{Ei}(itE_1+\gamma t)-\mathrm{Ei}(itE_0+\gamma t)\right]$. Note that for for small $t$ one can employ the expansion of $\mathrm{Im}\left[\mathrm{Ei}(t(\gamma +iE_1))\right]=\mathrm{Im}\left[\gamma_E+\ln t+\ln(\gamma +i E_1)+O(t)\right]=\arctan\left(E_1/\gamma\right)$, where $\gamma_E$ is the Euler-Mascheroni constant. From here, the leading order expansion of the last line of Eq.~\eqref{eq: PAB} becomes $\mathrm{Im}\left[\mathrm{Ei}(t(\gamma +iE_1))-\mathrm{Ei}(t(-\gamma +iE_1))+\mathrm{Ei}(t(-\gamma +iE_0))-\mathrm{Ei}(t(\gamma +iE_0))\right]=2\left(\arctan(E_1/\gamma)-\arctan(E_0/\gamma)\right)$, cancelling the first term and so satisfying that $P(B,0\vert A)=0$

\begin{figure}%[htbp]
\includegraphics[width=0.45\linewidth]{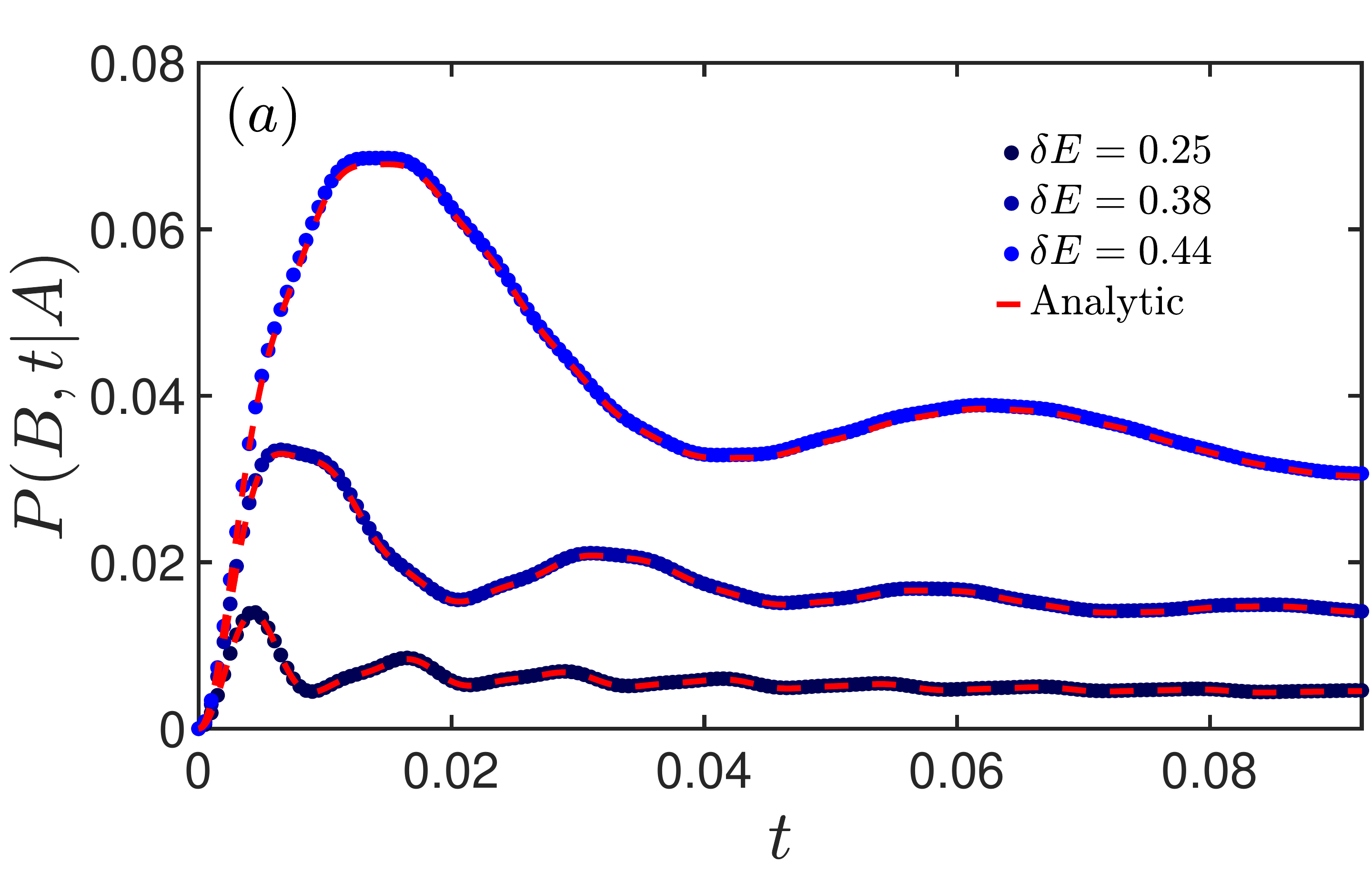}        
\includegraphics[width=0.45\linewidth]{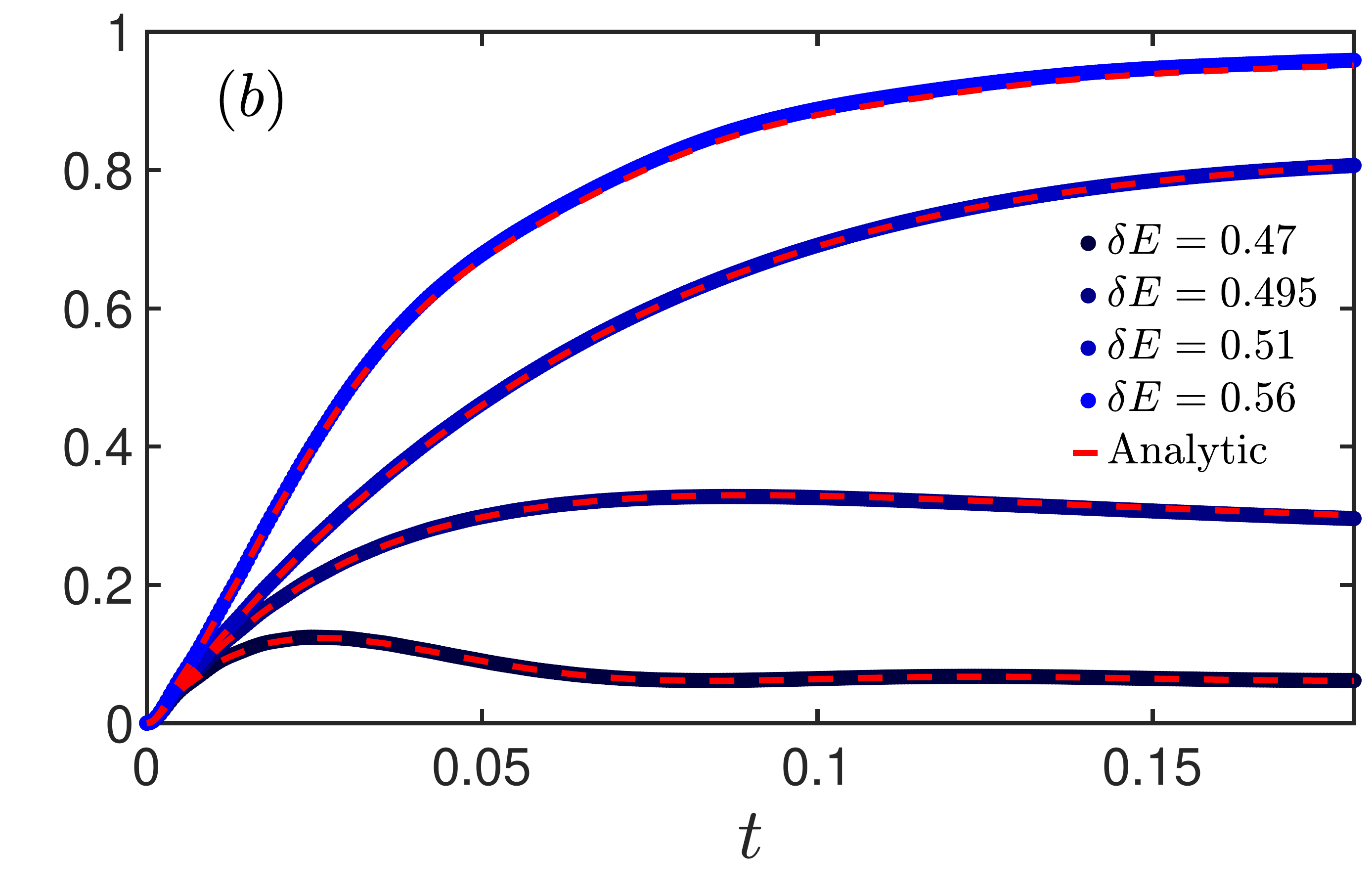}
\caption{Time-evolution of the transition probabilities into a target space of energy range $[E_0,E_1]$. 
(a) Strong suppression is observed for energy windows below zero energy, $E_1<0$, with a precise matching of the analytical results $(N=2000, W=2, \Delta =1)$. (b) For target subspaces involving also positive energies, the shape and order of magnitude of $P(B,T\vert A)$ changes dramatically, exhibiting a sigmoid-type shape and saturating to values close to unity.}
\label{fig:PAB_neg}
\end{figure}

The QTR also admits the closed form by using that $\mathrm d\mathrm{Ei}(t\gamma +itE_1)/\mathrm dt= e^{itE_1+t\gamma}/t$
\begin{eqnarray}\label{eq: kAB}
    k_{A\rightarrow B}(t)    
    &=&-2\frac{W^2}{\Delta}e^{-\frac{2\pi W^2t}{\Delta}}\left[\arctan\left(\frac{E_1\Delta}{\pi W^2}\right)-\arctan\left(\frac{E_0\Delta}{\pi W^2}\right)\right]\\
    &&+\frac{2W^2}{\Delta}\mathrm{Im}\Big[e^{-2\frac{\pi W^2}{\Delta}t}\left[\mathrm{Ei}(itE_1+t\pi W^2/\Delta )-\mathrm{Ei}(itE_0+t\pi W^2/\Delta)\right]\Big].\nonumber
\end{eqnarray}
Figure \ref{fig:PAB_neg} illustrates the conditional probability $P(B,t|A)$ for $W=2$, $N=2000$, and varying the target subspace.  The analytical result in Eq.~\eqref{eq: PAB} precisely captures the shape of $P(B,t\vert A)$, exhibiting characteristic suppression when the target subspace is confined to negative energies. At the same time, it changes significantly once the zero-energy line is crossed, showing a transition to a saturating curve.

\begin{figure}%[htbp]
\includegraphics[trim=0 0 0cm 0cm, clip,width=0.45\linewidth]{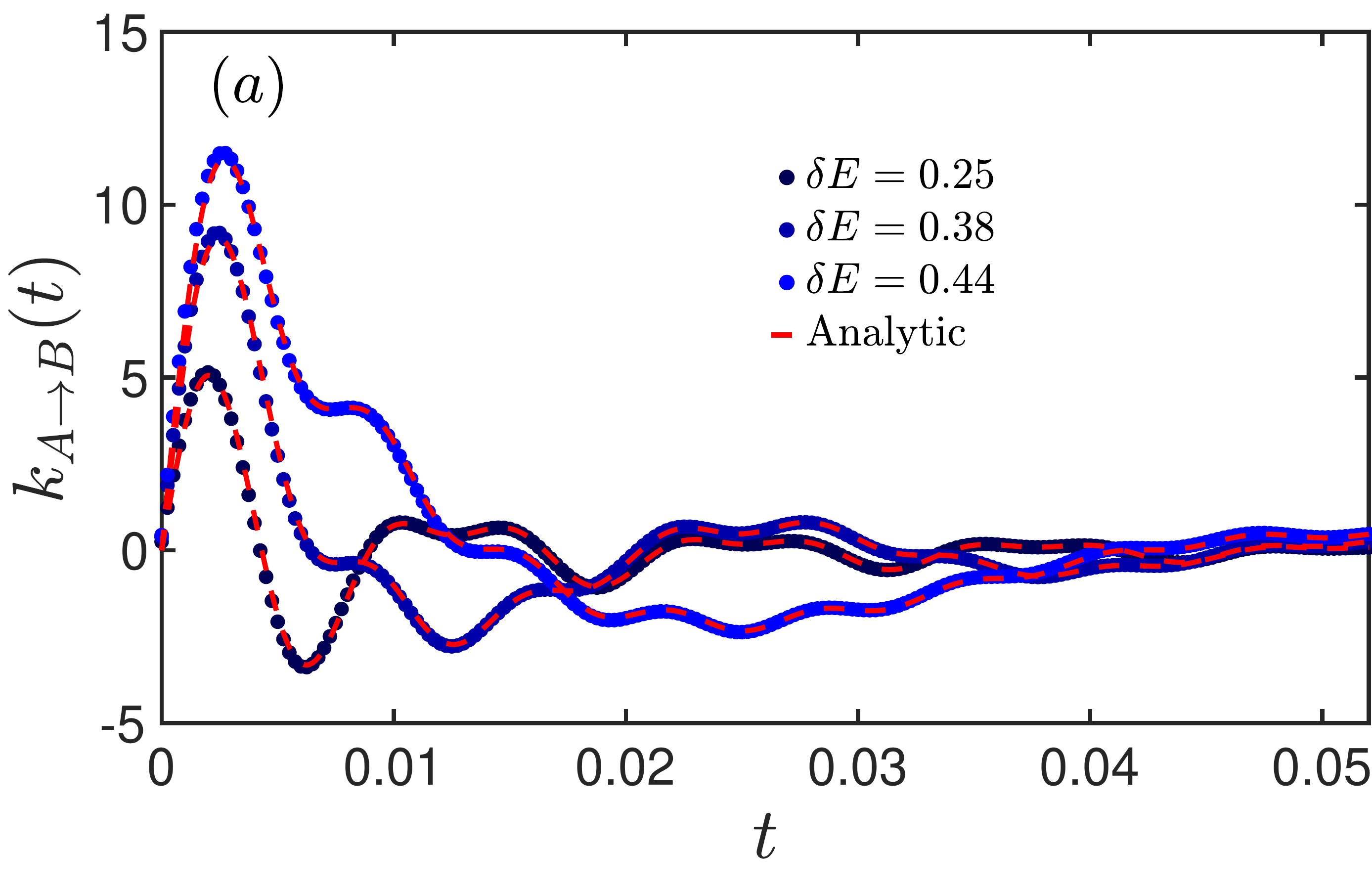}
\includegraphics[trim=1cm 0 0 0, clip,width=0.45\linewidth]{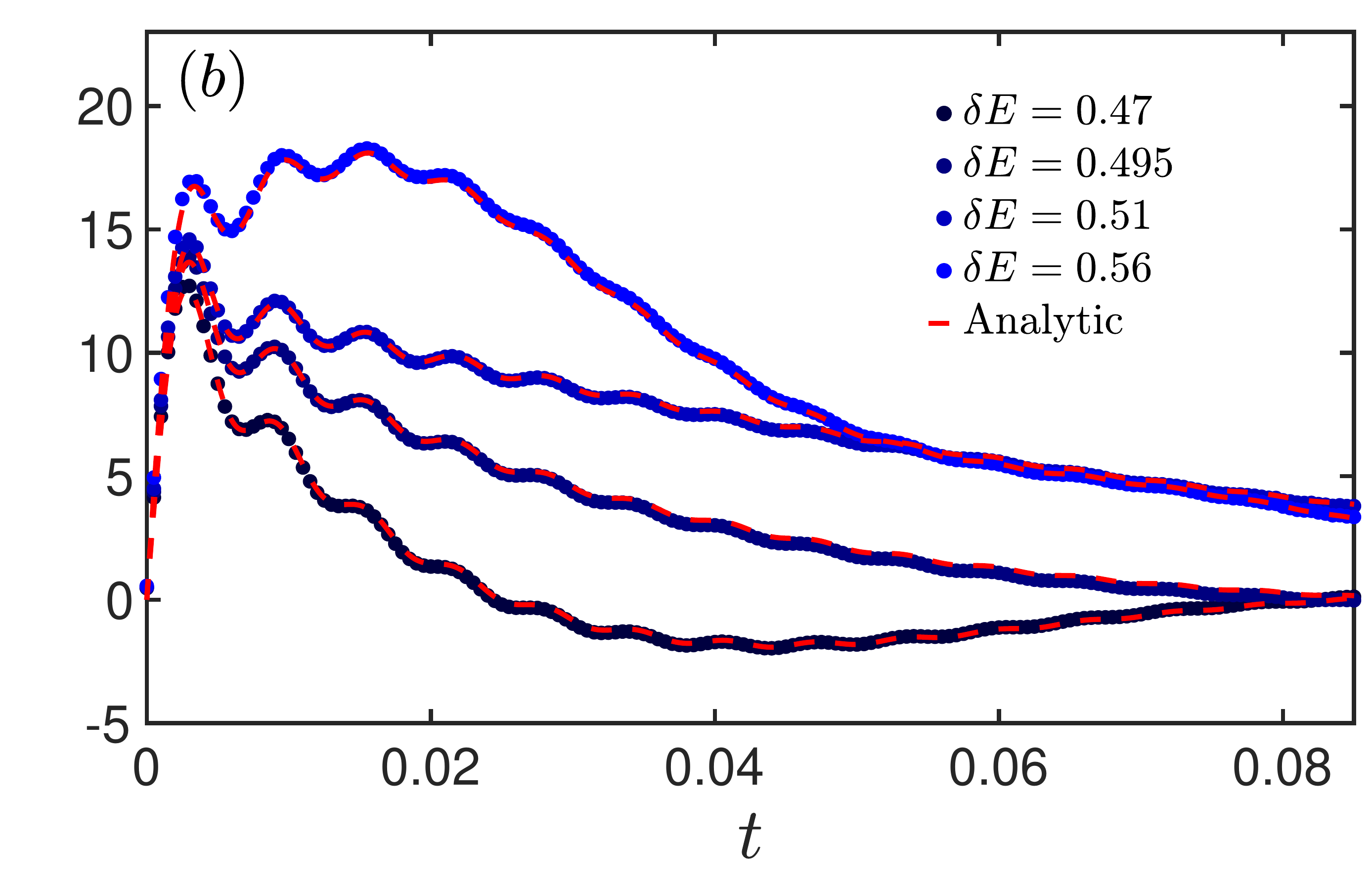}        
\caption{(a) QTRs for target subspace confined to negative energies exhibiting short-time peak in accordance with the transition probabilities. (b) For target subspace entering the positive energy region, a broader QTR traces out in accordance with the sigmoid shape of the transition probabilities.}
\label{fig:kAB}
\end{figure}

The corresponding QTRs are shown in Fig.~\ref{fig:kAB}, showing perfect agreement with the analytical results, Eq.~\eqref{eq: kAB}. For target subspaces confined to the negative energies, changes in $k_{A\rightarrow B}(t)$ are more pronounced at short times, whereas they extend to longer time scales when the target subspace enters the positive energy part. Note that QTRs need not be positive, and can take negative values as shown for $\delta E=0.47$. QTRs can thus be used to quantify the backflow effect \cite{TQM1,TQM2}. 

\section{Quantum speed limit in the Bixon-Jortner model}
In this section, we further demonstrate the tightness of the QTR-based QSL in the Bixon-Jortner model. Using the exact result in Eq.~\eqref{eq: PAB} and the condition in Eq.~\eqref{eq: P_AB_P_AA} of the main text for the initial pure state case, one obtains for a given energy window, $[E_0,E_1]$,
\begin{eqnarray}
    &&-\frac{1}{\pi}\mathrm{Im}\left[e^{-\frac{2\pi W^2}{\Delta}t}\left[\mathrm{Ei}(itE_1+t\pi W^2/\Delta )-\mathrm{Ei}(itE_0+t\pi W^2/\Delta)\right]-\left[\mathrm{Ei}(itE_1-t\pi W^2/\Delta)-\mathrm{Ei}(itE_0-t\pi W^2/\Delta)\right]\right]\nonumber\\
    &&+\frac{1+e^{-\frac{2\pi W^2t}{\Delta}}}{\pi}\left[\arctan\left(\frac{E_1\Delta}{\pi W^2}\right)-\arctan\left(\frac{E_0\Delta}{\pi W^2}\right)\right]\leq \frac{E_1-E_0}{\Delta}e^{-\frac{2\pi W^2 t}{\Delta}}.\nonumber\\
    &&\Rightarrow e^{\frac{2\pi W^2 t}{\Delta}}\left[\arctan\left(\frac{E_1\Delta}{\pi W^2}\right)-\arctan\left(\frac{E_0\Delta}{\pi W^2}\right)\right]\leq \frac{E_1-E_0}{\Delta}\pi\Rightarrow e^{\frac{2\pi W^2t}{\Delta}}\leq \frac{E_1-E_0}{\Delta\left[\arctan\left(\frac{E_1\Delta}{\pi W^2}\right)-\arctan\left(\frac{E_0\Delta}{\pi W^2}\right)\right]}\pi,
\end{eqnarray}
where the dimensionality of the final density matrix is taken into account by the discretization of the model with the level spacing $\Delta$. Here, we also used the large $t$ leading order behavior of $\mathrm{Ei}(itE_1+t\gamma)\sim e^{itE_1-\gamma t}/t$ and can thus be neglected compared to the constant term.
Thus for a threshold time diverging logarithmically with the number of states,
\begin{eqnarray}
    t^*=\frac{\Delta}{2\pi W^2}\log\left[\frac{E_1-E_0}{\Delta}\right],
\end{eqnarray}
the QTR-based speed limit provides a tighter QSL. For the QSL, the variance of the energy in the initial state is given by
\begin{eqnarray}
    \Delta^2_{\hat\rho_A}\hat H=\langle 0\lvert H^2\rvert0\rangle=\sum_nW^2_n=\Delta  E W^2,
\end{eqnarray}
where $\Delta E$ is the total bandwidth. Thus, the corresponding QSL is given by
\begin{eqnarray}
    \tau_\mathrm{QTR}=\frac{\cos^{-1}\left[\sqrt{P(B,t\vert A)\Delta /(E_1-E_0)}\right]}{W\sqrt{\Delta E}},\quad t<t^*
\end{eqnarray}
with $P(B,t\vert A)$ given in Eq.~\eqref{eq: PAB} for $[E_0,E_1]$.

\section{Super fidelity bound and speed limit}
The Uhlmann fidelity can be upper-bounded using the super fidelity  bound \cite{Miszczak2008SubAS},
\begin{eqnarray}
    F(\tau)=F(\hat \rho_A(\tau),\hat \pi_B)=\tr(\sqrt{\sqrt{\hat \rho_A(\tau)}\hat \pi_B\sqrt{\rho_A(\tau)}})^2\leq\tr(\hat\rho_A(\tau)\hat\pi_B)+M_{\hat\rho_A(\tau)}M_{\hat\pi_B}=P(B,\tau\vert A)d^{-1}_B+M_{\hat\rho_A(\tau)}M_{\hat\pi_B},\nonumber\\
\end{eqnarray}
where $M_{\hat\rho_A(\tau)}=\sqrt{1-\tr(\hat\rho^2_A(\tau))}$ and $M_{\hat\pi_B}=\sqrt{1-\tr(\hat\pi^2_B)}$ denote the mixedness of the two density matrices of $\hat\rho_A$ and $\hat\pi_B$, respectively.
We note that this bound cannot be derived from the Fuchs van de Graaf inequality \cite{Bengtsson2017} and also provides a tighter bound than it.
Upon using it, the Mandelstam-Tamm bound naturally follows from the inequality between the Bures length and the quantum Fisher information $\mathcal L(\tau)=\cos^{-1}\left(\sqrt{F(\tau)}\right)\leq\frac{1}{\tau}\int_0^\tau\mathrm dt\Delta_{\rho_A(t)}H(t)$,
\begin{eqnarray}
    \tau\geq\frac{\hbar\mathcal L(\tau)}{\frac{1}{\tau}\int_0^\tau\mathrm dt\Delta_{\hat\rho_A(t)}\hat H(t)}\geq\frac{\hbar\cos^{-1}\left[\sqrt{P(B,\tau|A)d^{-1}_B+M_{\hat\rho_A(\tau)}M_{\hat\pi_B}}\right]}{\frac{1}{\tau}\int_0^\tau\mathrm dt\Delta_{\hat\rho_A(t)}\hat H(t)}.
\end{eqnarray}

In deriving these results, we have invoked the notion of the maximally mixed state $\hat \pi_B$, which weights equally each rank-one projector spanning the subspace $\mathcal{H}_B$.
Alternative constructions can be envisioned. For instance, whenever the projector associated with the target subspace admits the form $\hat\Pi_B=\sum_{j=1}^{d_B}|j\rangle\langle j|$, one can consider the set of Bures lengths $\mathcal{L}(\hat\rho_A(t),|j\rangle\langle j|)=\cos^{-1}\sqrt{\langle j|\hat\rho_A(t)|j\rangle}$ and find the shortest distance minimizing over $j$. Such an approach is closer in spirit to, e.g., the formulation of the time of arrival problem and chemical reaction rates, in which $j$ is generally considered as a continuous index.

\section{Quantum Zeno effect}

The dynamical map resulting from unitary evolution combined with projective measurements at time intervals $\delta t=t/n$ is written as $U_n(t)=\left(\hat\Pi\hat U(t/n)\hat\Pi\right)^n$ with an arbitrary $\hat\Pi$ projector. For intervals sufficiently smaller than any relevant timescale of the system, this becomes $U_n(t)\approx\lim_{n\rightarrow\infty} U_n=\exp\left[-it\hat\Pi\hat H\hat\Pi/\hbar\right]$ \cite{Facchi08}. The time-evolved projector, in turn, can be expressed as
\begin{eqnarray}
\hat\Pi_B(t)=\exp\left[-it\hat\Pi\hat H\hat\Pi/\hbar\right]\hat\Pi_B\exp\left[it\hat\Pi\hat H\hat\Pi/\hbar\right]= \lim_{n\rightarrow\infty}  \left[1-i\frac{t}{n}\hat\Pi\hat H\hat\Pi/\hbar\right]^n\hat\Pi_B\left[1+i\frac{t}{n}\hat\Pi\hat H\hat\Pi/\hbar\right]^n.
\end{eqnarray}

For $\hat\Pi=\hat\Pi_B$ and using that $\left[\hat\Pi_B,\hat\Pi_B\hat H\hat\Pi_B\right]=0$, one finds
\begin{eqnarray}
    \left[1-i\frac{t}{n\hbar}\hat\Pi_B\hat H\hat\Pi_B\right]\hat\Pi_B\left[1+i\frac{t}{n\hbar}\hat\Pi_B\hat H\hat\Pi_B\right]=\hat\Pi_B+\frac{it}{n\hbar}\left[\hat\Pi_B,\hat\Pi_B\hat H\hat\Pi_B\right]+O(t^2/n^2),
\end{eqnarray}
where the second term identically vanishes.
Thus, the final result up to the leading-order correction is given by
\begin{eqnarray}
    \hat \Pi_B(t)=\lim_{n\rightarrow\infty}\left(\hat\Pi_B+O(t^2/n^2)\right)^n=\hat\Pi_B.
\end{eqnarray}
where the leading-order correction is suppressed as $O(1/n)$.

Next, we show how to arrive at the leading-order expression for the transition probability in the time-independent case. Consider the short-time power series expansion of the time evolution operator. 
\begin{eqnarray}
    \hat U(t)=\mathbb I-i\frac{t}{\hbar}\hat H-\frac{t^2}{2\hbar^2}\hat H^2+O(t^3).
\end{eqnarray}
Using it, the time-evolved projector reads 
\begin{eqnarray}
    \hat \Pi_B(t)\simeq \hat\Pi_B+i\frac{t}{\hbar}[\hat\Pi_B,\hat H]+\frac{t^2}{\hbar^2}\left(\hat{H}\hat{\Pi}_B \hat{H}-\frac{1}{2}\left\{\hat{H}^2,\hat{\Pi}_B\right\}\right)+O(t^3).
\end{eqnarray}
Taking the trace with $\hat\rho_A$, only the third term survives as the subspaces $A$ and $B$ are disjoint, $\hat\rho_A\hat\Pi_B=0$. As a result, the leading order expansion of the transition probability reads
\begin{eqnarray}
    P(B,t\vert A)=\frac{t^2}{\hbar^2}\tr[\hat{\rho}_A\hat{H}\hat{\Pi}_B \hat{H}]+\mathcal{O}(t^3).
    \end{eqnarray}
Using this, the leading-order behavior of the QTR is given by
\begin{eqnarray}
    k_{A\rightarrow B}=\frac{2t}{\hbar^2}\tr[\hat{\rho}_A\hat{H}\hat{\Pi}_B \hat{H}]+O(t^2).
\end{eqnarray}
For sufficiently small intervals $\delta t$, the transition rate can be written as
$k_{A\rightarrow B}\approx P(\delta t)/\delta t$. The condition for negligible transition probabilities $\delta t\ll \hbar/\tr[\hat{\rho}_A\hat{H}\hat{\Pi}_B \hat{H}]^{1/2}$ implies that$k_{A\rightarrow B}(\delta t)\ll \delta t^{-1}$.
%, leading to the same bound on $\delta t$.

One can also generalize this conclusion to driven quantum systems. 
Using the Dyson series for the time evolution operator generated by $\hat{H}(t)$,
\begin{eqnarray}
\hat{U}(t,0)=\mathbb{I}+\sum_{n=1}^\infty\left(\frac{1}{i\hbar}\right)^n \int_0^tdt_1\int_0^{t_1}dt_2\cdots\int_{0}^{t_{n-1}}dt_n\hat{H}(t_1)\hat{H}(t_2)\cdots \hat{H}(t_n).
\end{eqnarray}
The time-evolved projector expanded as
\begin{eqnarray}
    \hat \Pi_B(t)\simeq \hat\Pi_B+\frac{i}{\hbar}\int_0^t\mathrm dt_1[\hat\Pi_B,\hat H(t_1)]+\frac{1}{\hbar^2}\int_0^t\mathrm dt_1\int_0^t\mathrm dt_2\hat{H}(t_1)\hat{\Pi}_B \hat{H}(t_2)-\frac{1}{2\hbar^2}\int_0^t\int_0^{t_1}\mathrm dt_1\mathrm dt_2\left\{\hat{\Pi}_B,\hat{H}(t_1)\hat H(t_2)\right\}.\quad
\end{eqnarray}
Taking the trace with $\hat\rho_A$, only the third term contributes to $P(B,t|A)$, leading to the result in the main text, Eq.~\eqref{eq:Zeno_PAB}.

\section{Speed limit for the rate of change of the QTRs in the time-dependent case}
In the time-dependent case, the rate of change of the QTR can be upper-bounded as
\begin{eqnarray}
  \hbar  \left\lvert\frac{\mathrm d}{\mathrm dt}k_{A\rightarrow B}(t)\right\rvert&&=\left\lvert\left\langle\left[\hat J_B(t),\hat H(t)\right]\right\rangle_{\hat\rho_A}+\left\langle\left[\hat\Pi_B(t),\dot{\hat H}(t)\right]\right\rangle_{\hat\rho_A}\right\rvert
  \nonumber\\&&\leq\left\lvert\left\langle\left[\hat J_B(t),\hat H(t)\right]\right\rangle_{\hat\rho_A}\right\rvert+\left\lvert\left\langle\left[\hat\Pi_B(t),\dot{\hat H}(t)\right]\right\rangle_{\hat\rho_A}\right\rvert\nonumber\\
    &&\leq2\left[\Delta_{\hat\rho_A}\hat J_B(t)\Delta_{\hat\rho_A}\hat H(t)+\Delta_{\hat\rho_A}\hat\Pi_B(t)\Delta_{\hat\rho_A}\dot{\hat H}(t)\right],
\end{eqnarray}
where in the last step the Robertson inequality has been applied for both terms.

\section{QTR for Liouvillian dynamics}
When the Liouville von Neumann equation reads
\begin{equation}
\frac{d}{dt}\hat{\rho}(t)=\lim_{dt\rightarrow 0}\frac{\mathcal{E}_{t+dt}(\hat{\rho}_0)-\mathcal{E}_t(\hat{\rho}_0)}{dt}=\mathcal{L}[\hat{\rho}(t)],    
\end{equation}
the quantum flux is given by
\begin{equation}
\hat{J}_B \left( t \right) 
=\mathcal{L}^\dag [\hat{\Pi}_B(t)],
\end{equation}
and the QTR equals 
\begin{equation}
k_{A\rightarrow B}(t)=\tr\left(\hat{\rho}_A\mathcal{L}^\dag [\hat{\Pi}_B(t)]\right).
\end{equation}

Consider the decomposition of the Liouvillian $\mathcal{L}(\circ)=[\hat{H},\circ]/(i\hbar)+\mathcal{D}(\circ)$, involving the Hamiltonian (possibly including a Lamb shift) and the dissipator $\mathcal{D}$. It follows that the quantum flux has the following contributions, one Hamiltonian and the second dissipative: $\hat{J}_B(t)=\frac{1}{i\hbar}[\hat{\Pi}_B(t),\hat{H}]+\mathcal{D}^\dag [\hat{\Pi}_B(t)]$. In the specific case of Markovian dynamics, the dissipator can be written in the Lindblad form and the adjoint Lindblad generator acting on an observable $O$ reads $\mathcal{L}^\dag [\hat{O}(t)]=i[\hat{H},\hat{O}(t)]/\hbar+\sum_n\gamma_n(\hat{L}_n^\dag 
\hat{O} \left( t \right) 
%O(t)
\hat{L}_n-\{\hat{L}_n^\dag 
\hat{L}_n,
\hat{O} \left( t \right) 
%O(t)
\}/2)$. 

\section{QTR using the operator sum representation}
Any quantum channel admits a (Kraus) operator-sum representation. The dynamics can be derived from the unitary evolution of the system coupled to an environment. Let $\{|e_k\rangle\}$ denote a complete basis in the Hilbert space of the environment, and $\hat{U}_{SE}(t)$ describe the global time-evolution operator.  Choosing the initial state of the environment in a purified state $|e_0\rangle\langle e_0|$, $\hat{\rho}_t=\mathcal{E}_t(\hat \rho_0)=\sum_k\langle e_k|\hat{U}_{SE}(t)\hat{\rho}_0\otimes|e_0\rangle\langle e_0|\hat{U}^\dag_{SE}(t)|e_k\rangle=\sum_k\hat{M}_k(t)\hat{\rho}_0\hat{M}^\dag_k(t)$ with $\hat{M}_k(t)=\langle e_k|\hat{U}_{SE}(t)|e_0\rangle$. The adjoint quantum channel reads $\mathcal{E}_t^\dag(\rho_0)=\sum_k\hat{M}^\dag_k(t)\rho_0\hat{M}_k(t)$ and the QTR is given by
\begin{eqnarray}
k_{A\rightarrow B}(t)=2{\rm Re}\sum_k\tr[\hat{\rho}_A\dot{\hat{M}}^\dag_k(t)\hat{\Pi}_B \hat{M}_k(t)]. 
\end{eqnarray}

\section{QTRs under counterdiabatic driving: general case}
In this section, we provide essential details for deriving the bound for the QTR under CD without the assumption of $\left[\rho_A,H_0(0)\right]=0$. By inserting the identity in the trace, $\tr(\hat\rho_A\hat J_B(t))$, the general expression for the transition rate in the basis of $H_0(t)$ can be written as
\begin{eqnarray}
    k_{A\rightarrow B}(t)=\sum_{n,m}\langle m_0\lvert\hat\rho_A\rvert n_0\rangle\langle n_0\lvert\hat J_B(t)\rvert m_0\rangle.
\end{eqnarray}
Applying now the triangular and Heisenberg inequality for an off-diagonal matrix element in energy basis, $\left\lvert\left\langle n_0\left\lvert\left[\hat H_\mathrm{CD}(t),\hat J_B(t)\right]\right\rvert m_0\right\rangle\right\rvert\leq\left[\Delta_{\vert n_0\rangle}\hat H_\mathrm{CD}(t)\Delta_{\vert m_0\rangle}\hat\Pi_B+\Delta_{\vert m_0\rangle}\hat H_\mathrm{CD}(t)\Delta_{\vert n_0\rangle}\hat\Pi_B(t)\right]$, one finds the upper bound for the absolute value of the transition rate,
\begin{eqnarray}
    \lvert k_{A\rightarrow B}(t)\rvert&&\leq\frac{1}{\hbar}\sum_{n,m}\lvert\langle m_0\lvert\hat\rho_A\rvert n_0\rangle\rvert\left[\Delta_{\vert n_t\rangle}\hat H_\mathrm{CD}\Delta_{\vert m_t\rangle}\hat\Pi_B+\Delta_{\vert m_t\rangle}\hat H_\mathrm{CD}\Delta_{\vert n_t\rangle}\hat\Pi_B\right]\\
    &&=\sum_{n,m}\lvert\langle m_0\lvert\hat\rho_A\rvert n_0\rangle\rvert\left[\sqrt{g^{(n)}_{\mu\nu}(t)\dot\lambda^\mu\dot\lambda^\nu}\Delta_{\vert m_t\rangle}\hat\Pi_B+\sqrt{g^{(m)}_{\mu\nu}(t)\dot\lambda^\mu\dot\lambda^\nu}\Delta_{\vert n_t\rangle}\hat\Pi_B\right],
\end{eqnarray}
where we switched back to the %Schr\"odigner
Schr\"odinger 
picture and used again the relation, $\hbar^2\Delta_{|n_t\rangle}\hat H_{\rm CD}^2=g^{(n)}_{\mu\nu}(t)\dot\lambda^\mu\dot\lambda^\nu$.

As the absolute value of the density matrix element is symmetric by using the simple inequality of $\Delta_{\vert m_t\rangle}\hat\Pi^2_B=\langle m_t\lvert\hat\Pi_B\rvert m_t\rangle\left(1-\langle m_t\lvert\hat\Pi_B\rvert m_t\rangle\right)\leq1/4$. Alternatively, one can use the semi-norm inequality \cite{Boixo07} along with the fact that the eigenvalues of the projector are $\{0,1\}$. 
The bound then simplifies to
\begin{eqnarray}
    \lvert k_{A\rightarrow B}(t)\rvert&&\leq\frac{1}{2}\sum_{n,m}\lvert\langle m_0\lvert\hat\rho_A\rvert n_0\rangle\rvert\sqrt{g^{(n)}_{\mu\nu}(t)\dot\lambda^\mu\dot\lambda^\nu},
\end{eqnarray}
which is a quantum geometric quantity. Specifically, for a monotonic driving of the control parameters $\lambda^\mu$,  the QTR upper bound is independent of the specific modulation in time. 
In addition, under the minimal assumption of $\left[\hat\rho_A,\hat H_0(0)\right]=0$, the double sum simplifies to a single one, by orthogonality, $\langle m_0\lvert\hat\rho_A\rvert n_0\rangle=\delta_{nm}\langle n_0\lvert\hat\rho_A\rvert n_0\rangle$
\begin{eqnarray}
    \lvert k_{A\rightarrow B}(t)\rvert&&\leq\frac{1}{2}\sum_{n}\lvert\langle n_0\lvert\hat\rho_A\rvert n_0\rangle\rvert\sqrt{g^{(n)}_{\mu\nu}(t)\dot\lambda^\mu\dot\lambda^\nu}.
\end{eqnarray}

Finally, we also provide upper bounds on the transition probability by using the triangle inequality $P(B,t|A)\leq \int_0^t dt'|k_{A\rightarrow B}(t')|$ as
\begin{equation}
P(B,t|A)\leq \frac{1}{2}\sum_{n,m}\lvert\langle n_0|\hat{\rho}_A|m_0\rangle\rvert \mathcal{L}(|m_0\rangle,|m_t\rangle),
\end{equation}
where we have identified the natural distance,  
$\mathcal{L}(|m_0\rangle,|m_t\rangle)=\int_0^tdt'\sqrt{g_{\mu\nu}^{(m)}(t')\dot{\lambda}^\mu\dot{\lambda}^\nu}=\int_{\lambda_0}^{\lambda_t}\sqrt{g_{\mu\nu}^{(m)}\mathrm d\lambda^{\mu}\mathrm d\lambda^\nu}$,
between the $m$-th eigenstate of the initial Hamiltonian and the corresponding eigenstate of the final Hamiltonian. 
The geometric bounds for $P(B,t|A)$ and $k_{A\rightarrow B}(t)$ are governed by the distance traveled by each eigenmode, reflecting the fact that CD generates parallel transport while suppressing transitions among different energy eigenstates.

\section{Flux-Flux correlator in the Transverse field Ising model}
In this section, we consider the transverse field Ising model (TFIM), described by the Hamiltonian~\cite{Zurek2005Dynamics,Dziarmaga2005Dynamics,Suzuki2012Quantum,Sachdev2011Quantum}
\begin{eqnarray}
    \hat{H}(t) = - \sum_{j=1}^L \left( \hat{\sigma}^x_j \hat{\sigma}^x_{j+1} + g(t) \hat{\sigma}_j^z \right),\quad g(t)=g(0)(1-t/\tau),
\end{eqnarray}
with the $\hat\sigma^{z,x}_j$ denoting the Pauli spin-$1/2$ operators at the $j$-th site.
By the Jordan-Wigner transformation of the spin operators, $\hat\sigma^x_j=2c^\dagger_jc_j-1$ and $\hat\sigma^z_j=\prod_{j=1}^{k-1}\hat\sigma^x_j(\hat c^\dagger_k+\hat c_k)/2$ with $\hat c^\dagger_j,\,\hat c_j$ denoting fermionic creation and annihilation operator and a subsequent Fourier decomposition of $\hat c_k=\frac{e^{-i\pi /4}}{\sqrt L}\sum_j e^{-i\pi jk}\hat c_j$ one arrives at a sum of independent two-level systems of 
\begin{eqnarray}
    \hat{H} = 2 \sum_{k>0} \hat{\psi}_k^\dagger \left[ (g(t)-\cos k)\hat\sigma^z_k + \sin k \, \hat\sigma^x_k \right] \hat{\psi}_k 
    = 2 \sum_{k>0} \hat{\psi}_k^\dagger H_k(t) \hat{\psi}_k,
\end{eqnarray}
where $\hat{\psi}_k := (\hat{c}_k, \hat{c}_{-k}^\dagger)^T$ is a vector of creation and annihilation operators for fermions of momentum $k$, and $\hat\sigma^{x,y,z}_k$ are another set of Pauli matrices, acting on the internal space of $\hat{\psi}_k$.
We exemplify the flux-flux correlator in the TFIM. The flux-flux correlation function for the the initial ground state, $\hat\rho_A$ and with the target subspace given by the final ground state up to a given cut-off momentum $k_B$ is given by
\begin{eqnarray}
    C_{AB}(t)=\tr[\hat\rho_A\hat J_B(t)\hat J_{A,t}],\,J_{B}(t)=(i\hbar)^{-1}\left[\hat\Pi_B(t),\hat H(t)\right],\,\hat J_{A,t}=(i\hbar)^{-1}\left[\hat\Pi_A,\hat H(t)\right],
\end{eqnarray}
with
\begin{eqnarray}
    &&\hat\rho_A=\prod_{k>0}\lvert 0\rangle_{k}\lvert 0\rangle_{-k}\langle 0\rvert_{-k}\langle 0\rvert_{k},\quad\\
    &&
    \hat\Pi_B=\prod_{0<k<k_B}\left(\sin\frac{k}{2}\lvert0\rangle_k\lvert0\rangle_{-k}-\cos\frac{k}{2}\lvert1\rangle_k\lvert1\rangle_{-k}\right)\left(\sin\frac{k}{2}\langle0\rvert_{-k}\langle0\rvert_k-\cos\frac{k}{2}\langle1\rvert_{-k}\langle1\rvert_k\right)\\
    &&=\prod_{k_B>k>0}\left[\sin^2\frac{k}{2}\lvert0\rangle_k\lvert0\rangle_{-k}\langle0\rvert_{-k}\langle0\rvert_k-\frac{\sin k}{2}\Big(\lvert1\rangle_k\lvert1\rangle_{-k}\langle0\rvert_{-k}\langle0\rvert_k+\lvert0\rangle_k\lvert0\rangle_{-k}\langle1\rvert_{-k}\langle1\rvert_k\Big)+\cos^2\frac{k}{2}\lvert1\rangle_k\lvert1\rangle_{-k}\langle1\rvert_{-k}\langle1\rvert_k\right],\nonumber
\end{eqnarray}
where $\lvert 0\rangle_k,\,\lvert 1\rangle_k$ are the eigenstates of the particle number operator $\hat c^\dagger_k\hat c_k$ of the $k$-th mode.
Here, using the fact that the $k$-th mode is evolved as $\hat U_k(t)\lvert0\rangle_k\lvert0\rangle_{-k}=u_k\lvert1\rangle_k\lvert1\rangle_{-k}+v_k\lvert0\rangle_k\lvert0\rangle_{-k}$ and $\hat U_k(t)\lvert1\rangle_k\lvert1\rangle_{-k}=v^*_k\lvert1\rangle_k\lvert1\rangle_{-k}-u^*_k\lvert0\rangle_k\lvert0\rangle_{-k}$, where $u_k$ and $v_k$ are the solutions to the time-dependent Schr\"odinger equation in the $k$-th mode. Altogether, one obtains for the time-evolved projector, $\hat\Pi_B(t)=\hat U^\dagger\hat\Pi_B\hat U$,
\begin{eqnarray}\label{eq: PI_B_TFIM}
    &&\hat\Pi_B(t)=\prod_{0<k<k_B}\left(u_{11}(k)\lvert0\rangle_k\lvert0\rangle_{-k}\langle0\rvert_{-k}\langle0\rvert_k+u_{12}(k)\lvert1\rangle_k\lvert1\rangle_{-k}\langle0\rvert_{-k}\langle0\rvert_k+u^*_{12}(k)\lvert0\rangle_k\lvert0\rangle_{-k}\langle1\rvert_{-k}\langle1\rvert_k+u_{22}(k)\lvert1\rangle_k\lvert1\rangle_{-k}\langle1\rvert_{-k}\langle1\rvert_k\right),\nonumber\\
    &&u_{11,k}(t)=\lvert v_k\rvert^2\sin^2\frac{k}{2}+(u_kv^*_k+v_ku^*_k)\frac{\sin  k}{2}+|u_k|^2\cos^2\frac{k}{2},\,u_{12,k}(t)=\frac{1}{2}\sin k\left((u^*_k)^2+v^2_k\right)-\cos k u^*_kv_k,\\
    &&u_{22,k}(t)=|u_k|^2\sin^2\frac{k}{2}-\frac{1}{2}\sin  k(u^*_kv_k+v^*_ku_k)+|v_k|^2\cos^2\frac{k}{2}.
\end{eqnarray}
The Hamiltonian in the Heisenberg picture is given by
\begin{eqnarray}
    &&\hat H(t)=\\
    &&2\sum_{k>0}\hat U^\dagger_k\Big[(g-\cos k)\Big(\lvert1\rangle_k\lvert1\rangle_{-k}\langle1\rvert_{-k}\langle1\rvert_k-\lvert0\rangle_k\lvert0\rangle_{-k}\langle0\rvert_{-k}\langle0\rvert_k\Big)+\sin k\Big(\lvert1\rangle_k\lvert1\rangle_{-k}\langle0\rvert_{-k}\langle0\rvert_k+\lvert0\rangle_k\lvert0\rangle_{-k}\langle1\rvert_{-k}\langle1\rvert_k\Big)\Big]\hat U_k\nonumber\\
    &&=h_{11,k}(t)\Big(\lvert1\rangle_k\lvert1\rangle_{-k}\langle1\rvert_{-k}\langle1\rvert_k-\lvert0\rangle_k\lvert0\rangle_{-k}\langle0\rvert_{-k}\langle0\rvert_k\Big)+h_{12,k}(t)\Big(\lvert1\rangle_k\lvert1\rangle_{-k}\langle0\rvert_{-k}\langle0\rvert_k+\lvert0\rangle_k\lvert0\rangle_{-k}\langle1\rvert_{-k}\langle1\rvert_k\Big),\nonumber\\
    &&h_{11,k}(t)=2(g-\cos k)\left(\lvert u_k\rvert^2-\lvert v_k\rvert^2\right)+4\sin k\mathrm{Re}(u^*_kv_k),\nonumber\\
    &&h_{12,k}(t)=-2(g-\cos k)(v_ku^*_k+v^*_ku_k) + \sin k\left(v^2_k+(v^*_k)^2-u^2_k-(u^*_k)^2\right).\nonumber
\end{eqnarray}
From this, the commutators in the correlation function are given by
\begin{eqnarray}
    &&\left[\hat\Pi_B(t),\hat H(t)\right]=\sum_{k>0}\left[-h_{12,k}\left(u_{11,k}-u_{22,k}\right)\right]\Big[\Big[\lvert1\rangle_k\lvert1\rangle_{-k}\langle0\rvert_{-k}\langle0\rvert_k-\lvert0\rangle_k\lvert0\rangle_{-k}\langle1\rvert_{-k}\langle1\rvert_k\Big]\\
    &&+2h_{11,k}u_{12,k}\lvert1\rangle_k\lvert1\rangle_{-k}\langle0\rvert_{-k}\langle0\rvert_k-2h_{11,k}u^*_{12,k}\lvert0\rangle_k\lvert0\rangle_{-k}\langle1\rvert_{-k}\langle1\rvert_k\Big]\prod_{0<q<k_B;q\neq k}\lvert 0\rangle_{q}\lvert 0\rangle_{-q}\langle 0\rvert_{-q}\langle 0\rvert_{q},\nonumber\\
    &&\left[\hat\Pi_A,\hat H(t)\right]=\sum_{k>0}h_{12,k}(t)\Big[\lvert1\rangle_k\lvert1\rangle_{-k}\langle0\rvert_{-k}\langle0\rvert_k-\lvert0\rangle_k\lvert0\rangle_{-k}\langle1\rvert_{-k}\langle1\rvert_k\Big]\prod_{0<q;q\neq k}\lvert 0\rangle_{q}\lvert 0\rangle_{-q}\langle 0\rvert_{-q}\langle 0\rvert_{q}.
\end{eqnarray}
As a result, the correlator becomes simply
\begin{eqnarray}
    C_{AB}(t)=\hbar^{-2}\sum_{k>0}h_{12,k}\left[4h_{11,k}\mathrm{Re}(u_{12,k})-h_{12,k}\left(u_{11,k}-u_{22,k}\right)\right].
\end{eqnarray}

\section{QSL in the TFIM for time-dependent driving}

Next, we consider the QSL in the TFIM. As the initial reduced density matrix is a pure state, the Uhlmann fidelity is proportional to the conditional probability,
\begin{eqnarray}
F(\hat\rho_A(\tau),\hat\pi_B)&=&\left(\tr[\sqrt{\sqrt{\hat\rho_A(\tau)}\hat\pi_B\sqrt{\hat\rho_A(\tau)}}]\right)^2=P(B,\tau\vert A)d^{-1}_B\nonumber\\
&\approx&d^{-1}_B\prod_{0<k<k_B}\left\lvert v_k\sin \frac{k}{2}-u_k\cos \frac{k}{2} \right\rvert^2\approx d^{-1}_Be^{-1.16\frac{L}{\sqrt{2\pi^3 \tau}}},\qquad
\end{eqnarray}
where the rank of the final subspace is given by the momentum, $d_B=2^{L(1-k_B/\pi)}$ and the Landau-Zener approximation was employed for the transition probabilities for slow driving, $\left\lvert v_k\sin \frac{k}{2}-u_k\cos \frac{k}{2} \right\rvert^2\approx 1-e^{-2\pi k^2\tau}$.
For the time average of the energy variance, we represent the Hamiltonian in Schr\"odinger picture. For the $k$-th mode it reads as
\begin{eqnarray}
    \hat H_S(t)&=&\sum_{k>0}\epsilon_k(t)\Big[\left(\cos\frac{\theta_k}{2}\lvert 0\rangle_{k,-k}+\sin\frac{\theta_k}{2}\lvert1\rangle_{k,-k}\right)\left(\cos\frac{\theta_k}{2}\langle 0\rvert_{k,-k}+\sin\frac{\theta_k}{2}\langle1\rvert_{k,-k}\right)\\
    &&-\left(\sin\frac{\theta_k}{2}\lvert 0\rangle_{k,-k}-\cos\frac{\theta_k}{2}\lvert1\rangle_{k,-k}\right)\left(\sin\frac{\theta_k}{2}\langle 0\rvert_{k,-k}-\cos\frac{\theta_k}{2}\langle1\rvert_{k,-k}\right)\Big]\nonumber\\
    &=&\sum_{k>0}\epsilon_k(t)\left[\cos\theta_k\left(\lvert 0\rangle\langle 0\rvert_{k,-k}-\lvert 1\rangle\langle 1\rvert_{k,-k}\right)+\sin\theta_k\left(\lvert 1\rangle\langle 0\rvert_{k,-k}+\lvert 0\rangle\langle 1\rvert_{k,-k}\right)\right],\nonumber
\end{eqnarray}
where we have employed the shorthand notation of $\lvert n\rangle_{k,-k}\equiv\lvert n\rangle_k\lvert n\rangle_{-k},\,n=0,1$.
The square of the Hamiltonian becomes,
\begin{eqnarray}
    \hat H^2_S(t)&=&\sum_{k>0}\epsilon^2_k(t)\Big[\left(\cos\frac{\theta_k}{2}\lvert 0\rangle_{k,-k}+\sin\frac{\theta_k}{2}\lvert1\rangle_{k,-k}\right)\left(\cos\frac{\theta_k}{2}\langle 0\rvert_{k,-k}+\sin\frac{\theta_k}{2}\langle1\rvert_{k,-k}\right)\\
    &&+\left(\sin\frac{\theta_k}{2}\lvert 0\rangle_{k,-k}-\cos\frac{\theta_k}{2}\lvert1\rangle_{k,-k}\right)\left(\sin\frac{\theta_k}{2}\langle 0\rvert_{k,-k}-\cos\frac{\theta_k}{2}\langle1\rvert_{k,-k}\right)\Big]\nonumber\\
    &&+2\sum_{k>k^\prime>0}\epsilon_{k}(t)\epsilon_{k^\prime}(t)\left[\cos\theta_k\left(\lvert 0\rangle\langle 0\rvert_{k,-k}-\lvert 1\rangle\langle 1\rvert_{k,-k}\right)+\sin\theta_k\left(\lvert 1\rangle\langle 0\rvert_{k,-k}+\lvert 0\rangle\langle 1\rvert_{k,-k}\right)\right]\nonumber\\
    &&\times\left[\cos\theta_{k^\prime}\left(\lvert 0\rangle\langle 0\rvert_{k^\prime,-k^\prime}-\lvert 1\rangle\langle 1\rvert_{k^\prime,-k^\prime}\right)+\sin\theta_{k^\prime}\left(\lvert 1\rangle\langle 0\rvert_{k^\prime,-k^\prime}+\lvert 0\rangle\langle 1\rvert_{k^\prime,-k^\prime}\right)\right]\nonumber\\
    &=&\sum_{k>0}\epsilon^2_k(t)\left[\lvert 0\rangle\langle 0\rvert_{k,-k}+\lvert 1\rangle\langle 1\rvert_{k,-k}\right]+2\sum_{k>k^\prime>0}\epsilon_{k}(t)\epsilon_{k^\prime}(t)\left[\cos\theta_k\left(\lvert 0\rangle\langle 0\rvert_{k,-k}-\lvert 1\rangle\langle 1\rvert_{k,-k}\right)+\sin\theta_k\left(\lvert 1\rangle\langle 0\rvert_{k,-k}+\lvert 0\rangle\langle 1\rvert_{k,-k}\right)\right]\nonumber\\
    &&\times\left[\cos\theta_{k^\prime}\left(\lvert 0\rangle\langle 0\rvert_{k^\prime,-k^\prime}-\lvert 1\rangle\langle 1\rvert_{k^\prime,-k^\prime}\right)+\sin\theta_{k^\prime}\left(\lvert 1\rangle\langle 0\rvert_{k^\prime,-k^\prime}+\lvert 0\rangle\langle 1\rvert_{k^\prime,-k^\prime}\right)\right].\nonumber
\end{eqnarray}
The time-evolved density matrix, on the other hand, gives
\begin{eqnarray}
    \hat\rho_A(t)=\prod_{k>0}\left[v_k(t)\lvert 0\rangle_{k,-k}+u_k(t)\lvert1\rangle_{k,-k}\right]\left[v^*_k(t)\langle 0\rvert_{k,-k}+u^*_k(t)\langle1\rvert_{k,-k}\right].
\end{eqnarray}
Thus, the trace average of the Hamiltonian and its square is given by
\begin{eqnarray}
    \langle \hat H_S(t)\rangle_{\hat\rho_A(t)}&=&\sum_{k>0}\epsilon_k(t)\left[\cos\theta_k\left(\lvert v_k(t)\rvert^2-\lvert u_k(t)\rvert^2\right)+2\sin\theta_k\mathrm{Re}\left\{u^*_k(t)v_k(t)\right\}\right]\\
    \langle \hat H^2_S(t)\rangle_{\hat\rho_A(t)}&=&\sum_{k>0}\epsilon^2_k(t)\left(\lvert v_k(t)\rvert^2+\lvert u_k(t)\rvert^2\right)+\\
    &&+2\sum_{k>k^\prime>0}\!\epsilon_k(t)\epsilon_{k^\prime}(t)\left[\cos\theta_k\left(\lvert v_k(t)\rvert^2-\lvert u_k(t)\rvert^2\right)+2\sin\theta_k\mathrm{Re}\left\{u^*_k(t)v_k(t)\right\}\right]\Big[\cos\theta_{k^\prime}\left(\lvert v_{k^\prime}(t)\rvert^2-\lvert u_{k^\prime}(t)\rvert^2\right)\nonumber\\
    &&+2\sin\theta_{k^\prime}\mathrm{Re}\left\{u^*_{k^\prime}(t)v_{k^\prime}(t)\right\}\Big]=\sum_{k>0}\epsilon^2_k(t)+\langle \hat H_S(t)\rangle_{\hat\rho_A(t)}^2-\sum_{k>0}\langle \hat H_{S,k}(t)\rangle^2_{\hat\rho_A(t)},\nonumber
\end{eqnarray}
where $\hat H_{S,k}(t)$ denotes the $k$-th Hamiltonian in Schr\"odinger picture.
Thus, as expected, the total variance simplifies to the sum of the TLS variances
\begin{eqnarray}
    \Delta^2_{\hat\rho_A(t)}\hat H_S(t)&=&\sum_{k>0}\Delta^2_{\hat\rho_A(t)}H_{S,k}=\sum_{k>0}\epsilon^2_k(t)-\langle \hat H_{S,k}(t)\rangle^2_{\hat\rho_A(t)}\\
    &=&\sum_{k>0}\epsilon^2_k(t)\Big[1-\cos^2\theta_k\left(\lvert v_k\rvert^2-\lvert u_k\rvert^2\right)^2-4\sin^2\theta_k\mathrm{Re}\left\{u^*_k(t)v_k(t)\right\}^2\nonumber\\
    &&-4\sin\theta_k\cos\theta_k\left(\lvert v_k\rvert^2-\lvert u_k\rvert^2\right)\mathrm{Re}\left\{u^*_k(t)v_k(t)\right\}\Big].\nonumber
\end{eqnarray}
As the time integral can not be expressed in a closed formula, we approximate it by the final value at $t=\tau$, with $\lvert v_k\rvert^2\approx e^{-2\pi \tau k^2}\equiv p_k,\,\lvert u_k\rvert^2\approx1-p_k,\,u^*_kv_k=\sqrt{p_k(1-p_k)}e^{i\varphi_k(\tau)},\,\varphi_k(\tau)=3\pi/4-2\tau\cos^2_k-2\tau\sin^2k\log(2\sqrt\tau\cos k)-\mathrm{arg}\Gamma(1+i\tau\sin^2k)$,  $\,\epsilon_k(\tau)=2$, $\theta_k(\tau)=k$.

Performing the momentum sums in the continuum limit, one finds
\begin{eqnarray}
    \sum_{k>0}\langle \hat H_{S,k}(t)\rangle^2_{\hat\rho_A(t)}&=&4\sum_k\cos^2 k\left(1-2p_k\right)^2+4\sin^2 k(p_k(1-p_k))\cos^2\varphi_k(\tau)+4\sin k\cos k\sqrt{p_k(1-p_k)}(1-2p_k)\cos\varphi_k(\tau)\nonumber\\
    &\approx&\frac{2L}{\pi}\int_0^\pi\mathrm dk\cos^2 k(1-2p_k)^2=L\left(1-\frac{\sqrt8-2}{\sqrt{\pi\tau}}\right),
\end{eqnarray}
where every term has been expanded up to the leading order in $k$ as the momentum integral for the higher order terms would have yielded larger powers of $\tau^{-1/2}$.

\begin{figure}
    \centering
    \includegraphics[width=.5\columnwidth]{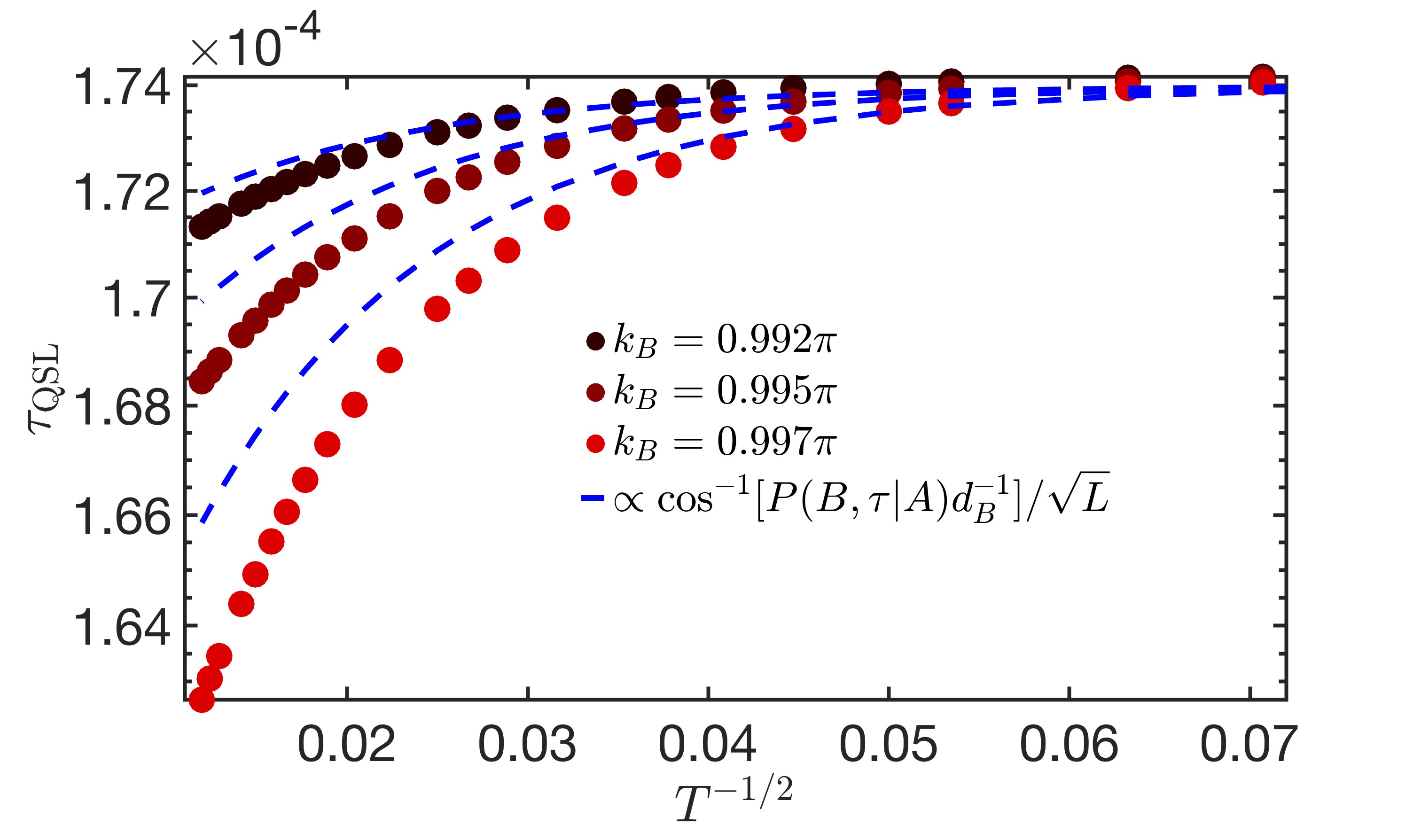}
    \caption{Quantum speed limits for different target momentum subspaces as a function of the final time. The inverse cosine function leaves room for almost no significant variations, but even at this scale, reasonable matching is found with the analytical approximations for the time-evolving energy variance  ($L=1000$).}\label{fig:QSL_TFIM} 
\end{figure}

Thus, combined with $\sum_{k>0}\epsilon^2_k(t)=2L$, the total variance is given by
\begin{eqnarray}
    &&\Delta^2_{\hat\rho_A(\tau)}\hat H_S(\tau)=L\left(1+\frac{\sqrt8-2}{\sqrt{\pi\tau}}\right).
\end{eqnarray}
According to numerical results, the variance monotonically decreases with $t$, thus the integral in the denominator of the speed limit can be lower bounded as
\begin{eqnarray}
    \frac{1}{\tau}\int_0^\tau\mathrm dt\,\Delta_{\hat\rho_A(t)}\hat H_S(t)\leq\Delta_{\hat\rho_A(\tau)}\hat H_S(\tau)\approx \sqrt L\sqrt{1+\frac{\sqrt8-2}{\sqrt{\pi\tau}}}.
\end{eqnarray}
As a result, the QSL in the time-dependent case can be estimated as
\begin{eqnarray}
\tau_\mathrm{QSL}\approx\frac{\hbar}{\sqrt L}\frac{\cos^{-1}\left[e^{-1.16\frac{L}{\sqrt{8\pi^3 \tau}}}2^{-\frac{L}{2}(1-k_B/\pi)}\right]}{\sqrt{1+\frac{\sqrt8-2}{\sqrt{\pi\tau}}}}.
\end{eqnarray}
The numerical verification of the above approximations are shown in Fig.~\ref{fig:QSL_TFIM}.

Next, we use the exact result to compare the condition of having a tighter QSL from the QTRs than from the MT bound. For this, we compute the survival probability, given by
\begin{eqnarray}
    P(A,\tau\vert A)=\tr[\hat\rho_A(\tau)\hat\Pi_A]=\prod_{k<k_B}\lvert v_k\rvert^2\approx e^{-\frac{L}{\pi}\int_0^{k_B}\mathrm dk \log(p_k)}\approx e^{-\frac{2L\tau}{3}k^3_B}.
\end{eqnarray}
where $\hat\Pi_A=\hat\rho_A(0)$ which only selected out the round state projectors for each $k$ mode from $\rho_A(\tau)$ in Eq.~\eqref{eq: rho_A(t)} and $p_k=e^{-2\pi \tau k^2}$. As to the leading order of $\tau$ the energy variance denominators are constants,
the condition for the QTR-based QSL being tighter is still captured by $P(B,t\vert A)<2^{L(1-k_B/\pi)}P(A,t\vert A)$, 
\begin{eqnarray}
    e^{-1.16 \frac{L}{\sqrt{2\pi^3 \tau}}}\leq 2^{L(1-k_B/\pi)}e^{-\frac{2L\tau}{3}k^3_B},
\end{eqnarray}
giving
\begin{eqnarray}
    \frac{2k^3_B\tau}{3}-\frac{1.16}{\sqrt{2\pi^3\tau}}+\frac{k_B}{\pi}\log 2\leq\log 2 ,
\end{eqnarray}
which, for small momentum $k_B\leq \tau^{-1/2}$, capturing the essential physics of the slow driving regime, translates to
\begin{eqnarray}
    \left(\frac{2}{3}-\frac{1.16}{\sqrt{2\pi^3}}+\frac{1}{\pi}\right)\tau^{-1/2}\leq \log 2\Rightarrow \tau\geq \left(\frac{2}{3}-\frac{1.16}{\sqrt{2\pi^3}}+\frac{1}{\pi}\right)^{2}(\log2)^{-2}\approx 1.46.
\end{eqnarray}
Thus, the condition is satisfied for driving rates covering the whole KZ scaling and adiabatic regimes, apart from a negligibly small regime close to the fast quench breakdown part. Furthermore, the threshold timescale falls properly within the KZ scaling regime, justifying the approximation of the slow driving limit, while the approximation of the variance also provides a satisfactory level of precision.

\section{QSL by QTRs under counterdiabatic driving in the TFIM}
In this section, we demonstrate the complementary speed limit approach obtained by bounding the transition rate in the presence of CD.
The transition rate with CD can be bounded as
\begin{eqnarray}
    \lvert k_{A\rightarrow B}(t)\rvert\leq\frac{1}{\hbar}\sum_n\langle n_0\lvert\hat\rho_A\rvert n_0\rangle\Delta_{\lvert n^{(M)}_t\rangle}\hat H_\mathrm{CD}\,\Delta_{\lvert n^{(M)}_t\rangle}\hat\Pi_B,
\end{eqnarray}
as the condition of $\left[\hat\rho_A,\hat H_0(0)\right]=0$ is satisfied for the same initial paramagnetic pure state as in Eq.~\eqref{eq: rho_A(t)}. Here, the states $\lvert n^{(M)}_t\rangle$ are evolved with the approximate CD of the $M$-th local expansion in the sudden quench limit,
\begin{eqnarray}
    \lvert n^{M}_{\mathrm{GS},t}\rangle\approx\prod_k\lvert\psi_{k,M}\rangle,\,\lvert\psi_{k,M}\rangle=\left[-\sin\left(\mathrm{Si}(Mk)-k/2\right),\cos\left(\mathrm{Si}(Mk)-k/2\right)\right]^T,
\end{eqnarray}
for large enough $M$~\cite{grabarits2025UniversalCD}.

The variance of the Hamiltonian in the sudden quench limit at the end of the evolution is simplified to $\Delta_{\lvert n^{(M)}_{\mathrm{GS},\tau}\rangle}\hat H_\mathrm{CD}=\langle n^{(M)}_{\mathrm{GS},\tau}\lvert\hat H^2_I\rvert n^{(M)}_{\mathrm{GS},\tau}\rangle$. The square of the CD term in the $M$-th order is given by 
\begin{eqnarray}
\hat H^2_I=\sum_{k>0}\hat\psi^\dagger_k\mathbb I_k\hat\psi_k\left(q^{(M)}_k(\tau)\right)^2,\quad q^{(M)}_k(\tau)\approx-\frac{1}{\tau}\frac{\sin k}{2}.
\end{eqnarray}
Here, the inverse of the driving times naturally emerges, also showing that the CD terms dominate all quantities over the bare Hamiltonian in the sudden limit,
\begin{eqnarray}
    \langle n^{(M)}_{\mathrm{GS},\tau}\lvert\hat H^2_I\rvert n^{(M)}_{\mathrm{GS},\tau}\rangle\approx\frac{1}{4\tau^2}\sum_{k>0}\sin^2k\approx \frac{L}{16\tau^2}.
\end{eqnarray}
The transition probability is expressed in a manner similar to that in Eq.~\eqref{eq: P_AB_CD}.
  Putting everything together, the bound can be written as
\begin{eqnarray}
    \lvert k_{A\rightarrow B}(\tau,M)\rvert\leq \sqrt{\langle n_{\mathrm{GS},T}\lvert\hat H^2_I\rvert n_{\mathrm{GS},\tau}\rangle
    P(B,\tau\vert A)\left[1-P(B,\tau\vert A)\right]}=\frac{\sqrt L}{4\tau}e^{-2.43\frac{L}{2\pi}M^{-1}}\sqrt{1-e^{-2.43\frac{L}{\pi}M^{-1}}}.
\end{eqnarray}
We also show how the conventional QTR-based and MT bounds can be obtained. First, we compute the time-average of the energy variance that reduces to the parametric integral of 
\begin{eqnarray}
\langle n^{(M)}_{\mathrm{GS},\tau}\lvert\hat H^2_I\rvert n^{(M)}_{\mathrm{GS},\tau}\rangle&=&\dot g^2\sum_k\sum_{m,m'=1}^M\frac{\sin(km)\sin(km')}{4}\frac{(g^{2m}+g^L)(g^{2m'}+g^L)}{g^{m+m'+2}(1+g^L)^2}\\
&=&\dot g^2\frac{L}{8}(1+g^L)^{-2}\sum_{m=1}^M\frac{(g^{2m}+g^L)^2}{g^{2m+2}}\\ &\approx& \dot g^2\frac{L}{8}
\begin{cases}
\sum_{m=1}^Mg^{2m-2}\approx(1-g^2)^{-1},\,g<1\\   
\sum_{m=1}^Mg^{-2m-2}\approx g^{-4}/(1-g^{-2}),\,g>1
\end{cases}\nonumber
\end{eqnarray}
where we have used that $\sum_k\sin(km)\sin(km')=L/2\,\delta_{mm'}$. Thus, the time integral becomes 
\begin{eqnarray}
    \frac{\sqrt L}{\sqrt 8}\left[\int_0^1\mathrm dg (1-g^2)^{-1/2}\,+\int_1^\infty\mathrm dgg^{-2}/\sqrt{1-g^{-2}}\right]=\sqrt{\frac{L}{8}}\pi,
\end{eqnarray}
becoming independent of the CD order for large $M$. Thus, again, the competition between the MT and the QTR-based QSLs will solely be determined by the transition and the survival probabilities. The latter is given by
\begin{eqnarray}
    P(A,\tau\vert A)&=&\prod_{k>0}\left\lvert\cos\left(\mathrm{Si}(Mk)-k/2\right)\right\rvert^2=\prod_{k>0}\left\lvert\cos\left(\mathrm{Si}(Mk)\cos(k/2)\right)+\sin\left(\mathrm{Si}(Mk)\sin(k/2)\right)\right\rvert^2\\
    &\approx&e^{\frac{L}{\pi}\int_0^\pi\mathrm dk\log\left[\cos\left(\mathrm{Si}(Mk)\cos(k/2)\right)+\sin\left(\mathrm{Si}(Mk)\sin(k/2)\right)\right]}\\
    &\approx& e^{-\frac{L}{\pi}M^{-1}\int_0^{Mk_B}\mathrm d\kappa\log\left[\cos\left(\mathrm{Si}(\kappa)\cos(\kappa/(2M))\right)+\sin\left(\mathrm{Si}(\kappa)\sin(\kappa/(2M))\right)\right]}\nonumber\\
    &\approx&e^{\frac{L}{\pi}M^{-1}\int_0^{Mk_B}\mathrm d\kappa\log\left[\cos\left(\mathrm{Si}(\kappa)\right) \right]}\approx e^{-\frac{L}{2\pi}M^{-1}\int_0^{Mk_B}\mathrm d\kappa\kappa^2}=e^{-\frac{L}{6\pi}M^2k^3_B},
\end{eqnarray}
where we used the small final momentum approximation, $k_B\leq 1/M$, capturing the essential physics for large enough locality order, $M$. As a result, the competition between the transition and survival probability governs the tightness of the QTR-based and MT speed limits in the leading order. Using the result of Eq.~\eqref{eq: P_AB_CD} for the transition probability, teh condition for a tighter QTR-based QSL reads as
\begin{eqnarray}
    e^{-2.48\frac{L}{\pi}M^{-1}}\leq 2^{L(1-k_B/\pi)} e^{-\frac{L}{6\pi}M^2k^3_B}&&\Rightarrow
    \frac{k^3_B}{6\pi}M^2-\frac{2.48}{\pi}M^{-1}+\frac{k_B}{\pi}\log2\leq \left(\frac{1}{6\pi}-\frac{2.48}{\pi}+\frac{1}{\pi}\log 2\right)M^{-1}\leq\log 2\nonumber\\
    &&\Rightarrow M\geq \left(\frac{1}{6\pi}-\frac{2.48}{\pi}+\frac{1}{\pi}\log 2\right)(\log 2)^{-1}\approx -0.744,
\end{eqnarray}
implying that for small momentum, whenever the large locality order is reached, the QTR-based QSL provides a tighter bound.

\section{Dirac delta potential in a box}
As a prominent example of the applicability of the QTRs between arbitrary abstract subspaces, we consider the particle-in-a-box model with a negative Dirac delta potential and between two infinite potential walls in the interval $[-a,a]$,, motivated by experimental realizations of trapped ions on various platforms~\cite{Gaunt_TrapBEC_2013,Navon2021_OpticalTrap}
\begin{eqnarray}\label{eq: H_Dirac}
    H=-\frac{1}{2}\partial^2_x+\gamma\delta(x),\,\gamma<0.
\end{eqnarray}
The corresponding eigenstates are given by
\begin{eqnarray}\label{eq: Dirac_box}
\psi_n^{\rm odd}(x) &=& \sqrt{\frac{1}{a}} \, \sin\left(\frac{n \pi x}{a}\right),\\
\psi_n^{\rm even}(x) &=& \sqrt{\frac{1}{N_n}} \, \cos(k_n x),\quad k_n \tan(k_n a) = \gamma,\,N_n=a+\frac{\sin(2k_na)}{2k_n}, \\
\psi_b(x) &=& \sqrt{\frac{\kappa}{\sinh(2 \kappa a) - 2 \kappa a}} \, \sinh\left(\kappa (a - |x|)\right),\quad  \kappa \coth(\kappa a) = |\gamma|,
\end{eqnarray}
where mass and the Planck constant have been set to unity.

We consider the transition rates from an initial subset $[m_1,m_2]$ of even states of the box potential to an orthonormal set $[n_1,n_2]$ such that $[n_1,n_2]\cap[m_1,m_2]=\emptyset$, set of even states with the time-evolution generated by the Hamiltonian in Eq.~\eqref{eq: H_Dirac}.
\begin{eqnarray}
    \psi_m(x,0)=\frac{1}{\sqrt a}\cos\left(\frac{(2m-1)\pi}{2a}x\right).
\end{eqnarray}
The evolution of this initial state is given in terms of the decomposition of the even states and the bound state (as the odd states are orthogonal),
\begin{eqnarray}
    \psi_m(x,t)&=&\sum_n c_{n,m}e^{-iE_n t}\psi^\mathrm{even}_n(x)+c_{b,m}e^{-i E_bt}\psi_b(x),\\
    c_{b,m}&=&\frac{1}{\sqrt{a}}\sqrt{\frac{\kappa}{\sinh(2 \kappa a) - 2 \kappa a}}\int_{-a}^{a} \, \sinh\left(\kappa (a - |y|)\right)\,\cos\left(\frac{(2m-1)\pi}{2a} y\right)\,dy,\quad \kappa \coth(\kappa a)=|\gamma|\\
    c_{n,m}&=&\frac{1}{\sqrt{a N_n}}\int_{-a}^{a}\cos(k_ny)\,\cos\left(\frac{(2m-1)\pi}{2a} y\right)\,dy,\quad k_n\tan(k_n a)=\gamma,\\
    c_{b,m}
&=&
\frac{1}{\sqrt{a}}
\sqrt{\frac{\kappa}{\sinh(2\kappa a)-2\kappa a}}\,
\frac{2\kappa\cosh(\kappa a)}{\kappa^2+\left(\frac{(2m-1)\pi}{2a}\right)^2},\\
c_{n,m}
&=&
\frac{1}{\sqrt{a\,N_n}}\,
\frac{2\left(\frac{(2m-1)\pi}{2a}\right)\cos(k_n a)(-1)^{m}}
{k_n^2-\left(\frac{(2m-1)\pi}{2a}\right)^2}.
\end{eqnarray}
Within this setup, the transition probability and QTRs read
\begin{eqnarray}
P([n_1,n_2],t \mid [m_1,m_2]) &=&
\sum_{m=m_1}^{m_2} \sum_{n=n_1}^{n_2} 
\Bigg| 
\sum_{n'} c_{n',m} \, e^{-i E_{n'} t} d_{n,n'} + c_{b,m} \, e^{-i E_b t} c_{b,n}
\Bigg|^2,\\
k([n_1,n_2],t \mid [m_1,m_2]) &=&
\sum_{m=m_1}^{m_2} \sum_{n=n_1}^{n_2} 2 \, \mathrm{Re} \Bigg\{
\Bigg( \sum_{n'} (-i E_{n'}) c_{n',m} e^{-i E_{n'} t} d_{n,n'} - i E_b c_{b,m} e^{-i E_b t} c_{b,n} \Bigg) \nonumber\\
&&\times \Bigg( \sum_{n'} c_{n',m}^* e^{i E_{n'} t} d_{n,n'}^* + c_{b,m}^* e^{i E_b t} c_{b,n}^* \Bigg) \Bigg\},
\end{eqnarray}
where the transition element between the even states arising in the transition probability is given by
\begin{eqnarray}
d_{n,n'}=\frac{1}{\sqrt{N_nN_{n'}}}\int_{-a}^a\mathrm dx\cos(k_nx)\cos(k_{n'}x)=\frac{2}{\sqrt{N_nN_{n'}}}\frac{\left(k_n\sin(k_{n'}a)\cos(k_na)-k_{n'}\sin(k_na)\cos(k_{n'}a)\right)}{k^2_n-k^2_{n'}}.
\end{eqnarray}
by exploiting that $k_n\tan(k_na)=\gamma.$
For the QSL, we restrict ourselves to the transition rate to a single even state and the bound state, starting from a single initial state. First, we need the time-average of the variance of the initial Hamiltonian, which is given by
\begin{eqnarray}
    \Delta H_m = \sqrt{
\sum_n |c_{n,m}|^2 \, \epsilon_n^2 + |c_{b,m}|^2 \, \epsilon_b^2
-
\left( \sum_n |c_{n,m}|^2 \, \epsilon_n - |c_{b,m}|^2 \, \epsilon_b \right)^2
}, \quad
\epsilon_n = \frac{k_n^2}{2}, \quad \epsilon_b = \frac{\kappa^2}{2}.
\end{eqnarray}
For rank-$1$ initial and target subspaces with $m\neq n$, the corresponding QTR-based speed limit is then given by
\begin{eqnarray}
\tau_\mathrm{QTR} =
\frac{
\hbar \, \cos^{-1} \Bigg[
\Bigg| 
\sum_{n'} c_{n',m} \, e^{-i E_{n'} t} \, d_{n,n'} + c_{b,m} \, e^{-i E_b t} \, c_{b,n}
\Bigg|
\Bigg]
}
{
\sqrt{
|c_{n,m}|^2 \, \epsilon_n^2 + |c_{b,m}|^2 \, \epsilon_b^2
-
\left( |c_{n,m}|^2 \, \epsilon_n - |c_{b,m}|^2 \, \epsilon_b \right)^2
}
}, \quad
\epsilon_n = \frac{k_n^2}{2}, \quad \epsilon_b = \frac{\kappa^2}{2}.
\end{eqnarray}
On the other hand, the MT depends on the initial fidelity based numerator with $P(m,t \mid m) = \Bigg| \sum_{n'} |c_{n',m}|^2 \, e^{-i E_{n'} t} + |c_{b,m}|^2 \, e^{-i E_b t} \Bigg|^2
$,
\begin{eqnarray}
\tau_\mathrm{MT}=\frac{\hbar\cos^{-1}\left[\Bigg| \sum_{n'} |c_{n',m}|^2 \, e^{-i E_{n'} t} + |c_{b,m}|^2 \, e^{-i E_b t} \Bigg|\right]}{\sqrt{\epsilon^2_n\lvert c_{n,m}\rvert^2+\epsilon^2_b\lvert c_{b,m}\rvert^2-\left(\epsilon_n\lvert c_{n,m}\rvert^2-\epsilon_b\lvert c_{b,m}\rvert^2\right)^2}}.
\end{eqnarray}
Given the condition of a better QTR-based bound, which can effectively be analyzed for $m=0$. Thus, to translate the inequality into a lower threshold $n$, 
\begin{eqnarray}
    \Bigg| 
\sum_{n'} c_{n',0} \, e^{-i E_{n'} t} \, d_{n,n'} + c_{b,0} \, e^{-i E_b t} \, c_{b,n}
\Bigg|\leq\Bigg| \sum_{n'} |c_{n',0}|^2 \, e^{-i E_{n'} t} + |c_{b,0}|^2 \, e^{-i E_b t} \Bigg|.
\end{eqnarray}

Using that $|d_{n,n'}|=\frac{2}{\sqrt{N_nN_{n'}}}\frac{|k_n\sin(k_{n'}a)\cos(k_na)-k_{n'}\sin(k_na)\cos(k_{n'}a)|}{k^2_n-k^2_{n'}}\leq \frac{2}{\sqrt{N_nN_{n'}}}\frac{1}{|k_n-k_{n'}|}$ and $|c_{n',0}|=\frac{2}{\sqrt{aN_{n'}}}\frac{k_0|\cos(k_na)|}{|k^2_{n'}-k^2_0|}\leq \frac{2k_0}{\sqrt{aN_{n'}}}\frac{1}{|k^2_{n'}-k^2_0|}$. Furthermore, $N_{n'}\geq a/2$, and $|k_n-k_{n'}|\geq\pi|n-n'|/a$, and $|k^2_n-k^2_0|\geq \pi^2(n'-1/2)(n'-3/2)/a^2$ one finds
\begin{eqnarray}
    |c_{n',0}d_{n,n'}|\leq\frac{\sqrt{128}k_0a}{\pi }\frac{1}{|n-n'||k^2_{n'}-k^2_0|}\leq \frac{16k_0a}{\pi^3}\frac{1}{|n-n'|(n'-1/2)(n'-3/2)},
\end{eqnarray}
by which

\begin{eqnarray}
    \Bigg| 
\sum_{n'} c_{n',0} \, e^{-i E_{n'} t} \, d_{n,n'} + c_{b,0} \, e^{-i E_b t} \, c_{b,n}
\Bigg|&\leq&\sum_{n'}\lvert c_{n',0}d_{n,n'}\rvert+\lvert c_{b,0}c_{b,n}\rvert\\
&\leq&\frac{16k_0a}{\pi^3
}\sum_{n'}\frac{1}{|n-n'|(n'-1/2)(n'-3/2)}+\lvert c_{b,0}c_{b,n}\rvert \nonumber\\
&\leq&\frac{16k_0a}{\pi^3}S_n+\frac{\kappa}{\sinh(2\kappa a)-2\kappa a}\frac{4\kappa^2\cosh^2(\kappa a)}{(\kappa^2+k^2_0)(\kappa^2+k^2_n)}\nonumber\\
&\leq&\frac{16k_0a}{\pi^3}S_n+\frac{4\kappa^3 a^2\cosh^2(\kappa a)}{\pi^2(\sinh(2\kappa a)-2\kappa a)(\kappa^2+k^2_0)}\frac{1}{(n-1/2)^2},\\
S_n&=&\sum_{n'\neq n}\frac{1}{|n-n'|(n'-1/2)(n'-3/2)}.
\end{eqnarray} 

The sum can be bounded as 
\begin{eqnarray}
    S_n&&=\sum_{m=1}^{n-2}\frac{1}{m(n-m-1/2)(n-m-3/2)}+\sum_{m=1}^\infty\frac{1}{m(n+m-1/2)(n+m-3/2)}\\
    &&\leq \sum_{m=1}^{n-2}\frac{4}{mn^2}+2\sum_{m=1}^\infty\frac{1}{m(n+m)^2}\leq 4\frac{\log n}{n^2}+\frac{1}{n^2}.\nonumber
\end{eqnarray}
Thus, a sane upper bound for the transition probability is given by 
\begin{eqnarray}
    P(n,t|0)\leq\left(\frac{16k_0a}{\pi^3}\frac{4\log n+1}{n^2}+\frac{4\kappa^3 a^2\cosh^2(\kappa a)}{\pi^2\sinh(2\kappa a)-2\kappa a)(\kappa^2+k^2_0)}\frac{1}{(n-1/2)^2}\right)^2.
\end{eqnarray}
Next, we lower bound the survival probability by using that $\sum_{n'}|c_{n',m}|^2+|c_{b,m}|^2=1$, and that the absolute value can be lower bounded by the worst case destructive interference scenario, one finds
\begin{eqnarray}
    \Bigg| \sum_{n'} |c_{n',0}|^2 \, e^{-i E_{n'} t} + |c_{b,0}|^2 \, e^{-i E_b t} \Bigg|\geq \left|1-2|c_{b,0}|^2\right|=\left\lvert1-2\frac{2\kappa}{\sinh(2\kappa a)-2\kappa a}\frac{4\kappa^2\cosh^2(\kappa a)}{(\kappa^2+(\pi/(2a))^2)^2}\right\rvert,
\end{eqnarray}
where we used that $k_0=\pi/(2a)$ for the zero mode.
Altogether one finds
\begin{eqnarray}
    &&\frac{16k_0a}{\pi^3}\frac{4\log n+1}{n^2}+\frac{4\kappa^3 a^2\cosh^2(\kappa a)}{\pi^2(\sinh(2\kappa a)-2\kappa a)(\kappa^2+k^2_0)}\frac{1}{(n-1/2)^2}\leq\left\lvert1-2\frac{2\kappa}{\sinh(2\kappa a)-2\kappa a}\frac{4\kappa^2\cosh^2(\kappa a)}{(\kappa^2+(\pi/(2a))^2)^2}\right\rvert\nonumber\\
    &&\Rightarrow n^*\equiv\left\lceil\frac{\frac{80k_0a}{\pi^3}+\frac{16\kappa^3a^2\cosh^2(\kappa a)}{\pi^2(\sinh(2\kappa a)-2\kappa a)(\kappa^2+k^2_0)}}{\left\lvert1-2\frac{2\kappa}{\sinh(2\kappa a)-2\kappa a}\frac{4\kappa^2\cosh^2(\kappa a)}{(\kappa^2+(\pi/(2a))^2)^2}\right\rvert}\right\rceil\leq \infty,
\end{eqnarray}
where in the last step we employed a generous upper bound as $(\log n+c)/n^2\leq (1+c)/n$. Here, $n^*$ is a finite integer, above which for infinitely many states the QTR-based QSL becomes tighter.

As a next step, we generalize this construction to a set of initial states, by which the desired inequality reads as 
\begin{eqnarray}
    \sum_{m\in[m_1,m_2]}\Bigg| 
\sum_{n'} c_{n',m} \, e^{-i E_{n'} t} \, d_{n,n'} + c_{b,m} \, e^{-i E_b t} \, c_{b,n}
\Bigg|^2\leq\sum_{m\in[m_1,m_2]}\Bigg| \sum_{n'} |c_{n',m}|^2 \, e^{-i E_{n'} t} + |c_{b,m}|^2 \, e^{-i E_b t} \Bigg|^2.
\end{eqnarray}
The survival probability can be lower-bounded by the overlaps with the bound state in the same spirit,
\begin{eqnarray}
    \sum_{m\in[m_1,m_2]}\Bigg| \sum_{n'} |c_{n',0}|^2 \, e^{-i E_{n'} t} + |c_{b,0}|^2 \, e^{-i E_b t} \Bigg|^2\geq\sum_{m\in[m_1,m_2]}\left|1-2c_{b,m}||^2\right|.
\end{eqnarray}
At the same time, for the transition probability, we apply similar techniques for each $m$,
\begin{eqnarray}
    \Bigg| 
\sum_{n'} c_{n',m} \, e^{-i E_{n'} t} \, d_{n,n'} + c_{b,m} \, e^{-i E_b t} \, c_{b,n}
\Bigg|\leq S^{(m)}_n+|c_{b,m}c_{b,n}|,\,S^{(m)}_n=\sum_{n'}|c_{n',m}d_{n,n'}|.
\end{eqnarray}
Using that $|c_{n',m}d_{n,n'}|\leq\frac{16\tilde k_ma}{\pi^3}\frac{1}{|n-n'|(n'-1/2)(n'-3/2)}$ all the previous techniques follow easily as
\begin{eqnarray}
    S^{(m)}_n\leq\frac{16\tilde k_ma}{\pi^3}\frac{4\log n+1}{n^2},
\end{eqnarray}
with $\tilde k_m=(2m-1)\pi/(2a)$ is the momentum for the box potential.
The bound state term can also be bounded as
\begin{eqnarray}
    |c_{b,m}c_{b,n}|\leq\frac{4\kappa^3\cosh^2(\kappa a)}{a\sinh(2\kappa a)-2\kappa a)}\frac{1}{(\kappa^2+k^2_m)(\kappa^2+k^2_n)}\leq\frac{16\kappa^3a\cosh^2(\kappa a)}{\pi^2(\sinh(2\kappa a)-2\kappa a)(\kappa^2+k^2_{m})}\frac{1}{(n-1/2)^2}.
\end{eqnarray}
As $\tilde k_m$ is an increasing function on $m$, one can upper bound the transition probability as
\begin{eqnarray}
    P(n,t|[m_1,m_2])&&\leq(m_2-m_1)\left[\frac{16\tilde k_{m_2}a}{\pi^3}\frac{4\log n+1}{n^2}+\frac{16\kappa^3a\cosh^2(\kappa a)}{\pi^2(\sinh(2\kappa a)-2\kappa a)(\kappa^2+k^2_{m_1})}\frac{1}{(n-1/2)^2}\right]\nonumber\\
    &&\leq (m_2-m_1)\left|1-2|c_{b,m_1}|^2\right|,
\end{eqnarray}
where in the numerator and in the denominator, the maximum and minimum $\tilde k_m=(2m-1)\pi/(2a)$ values were taken, respectively.
Further applying the upper bound $(4\log n+1)/n^2\leq 5/n$, one obtains a similar finite threshold $n^*$
\begin{eqnarray}
    n^*\equiv \left\lceil\frac{\frac{80\tilde k_{m_2}a}{\pi^3}+\frac{16\kappa^3a^2\cosh^2(\kappa a)}{\pi^2(\sinh(2\kappa a)-2\kappa a)(\kappa^2+k^2_{m_1})}}{\left\lvert1-2\frac{2\kappa}{\sinh(2\kappa a)-2\kappa a}\frac{4\kappa^2\cosh^2(\kappa a)}{(\kappa^2+k^2_{m_1})^2}\right\rvert}\right\rceil\leq \infty.
\end{eqnarray}
Finally, we show that the situation holds for arbitrary sets of initial and final states $[m_1,m_2]\cap[n_1,n_2]=\emptyset$, for which the desired inequality reads
\begin{eqnarray}
    \sum_{n\in[n_1,n_2]m\in[m_1,m_2]}\Bigg| 
\sum_{n'} c_{n',m} \, e^{-i E_{n'} t} \, d_{n,n'} + c_{b,m} \, e^{-i E_b t} \, c_{b,n}
\Bigg|^2&&+\sqrt{1-(m_2-m_1)^{-1}}\sqrt{1-(n_2-n_1)^{-1}}\\
&&\leq\sum_{m\in[m_1,m_2]}\Bigg| \sum_{n'} |c_{n',0}|^2 \, e^{-i E_{n'} t} + |c_{b,0}|^2 \, e^{-i E_b t} \Bigg|^2.\nonumber
\end{eqnarray}
where the square-root term encodes the mixednesses of the initial and target subspaces.
To this end, the same techniques can be applied for the survival probability, while the summation over the target states can be upper bounded by the same techniques, and taking the smallest $n_1$,
\begin{eqnarray}
    &&\frac{n_2-n_1}{n_1}(m_2-m_1)\left[\frac{80\tilde k_{m_2}a}{\pi^3}+\frac{16\kappa^3a^2\cosh^2(\kappa a)}{\pi^2(\sinh(2\kappa a)-2\kappa a)(\kappa^2+k^2_{m_1})}\right]+\sqrt{\left(1-(m_2-m_1)^{-1}\right)\left(1-(n_2-n_1)^{-1}\right)}\\
    &&\leq n_2(m_2-m_1)\left[\frac{80\tilde k_{m_2}a}{\pi^3}+\frac{16\kappa^3a^2\cosh^2(\kappa a)}{\pi^2(\sinh(2\kappa a)-2\kappa a)(\kappa^2+k^2_{m_1})}\right]+\sqrt{\left(1-(m_2-m_1)^{-1}\right)\left(1-(n_2-n_1)^{-1}\right)}\nonumber\\
    &&\leq (m_2-m_1)\left|1-2|c_{b,m_1}^2|\right|,\nonumber
\end{eqnarray}
where we have used that $(n_2-n_1)/n_1\leq n_2$. From here, the bound for $m_2<n_1$ reads
\begin{eqnarray}
    n^*_1\equiv n_2-\frac{1}{1-\frac{\left[(m_2-m_1)|1-2|c_{b,m_1}|^2|-n_2(m_2-m_1)\left(\frac{80 \tilde k_{m_2}a}{\pi^3}+\frac{16\kappa^3a^2\cosh^2(\kappa a)}{\pi^2\left(\sinh(2\kappa a)-2\kappa a\right)(\kappa^2+k^2_{m_1})}\right)\right]^2}{1-(m_2-m_1)^{-1}}}\leq \infty.
\end{eqnarray}
By this, again for arbitrary $n_2$ there is a finite lower bound for $n_1$ above which the QTR-based QSL is tighter than the MT bound.

\section{Relations to quenches in the TFIM and relaxation dynamics}
We consider the relation of the QTRs in the relaxation dynamics of the TFIM (for more details see Refs.~\cite{Calabrese_PRL2011,Calabrese_2012Review,Calabrese_2012_Review2}). To this end, we consider an initial state with a given $g_1$ magnetic field in the TFIM with a density matrix
\begin{eqnarray}
\hat{H}(t) &=& 2 \sum_{k>0} \hat{\psi}_k^\dagger \left[ (g(t)-\cos k)\, \hat\sigma^z_k + \sin k \, \hat\sigma^x_k \right] \hat{\psi}_k
= 2 \sum_{k>0} \hat{\psi}_k^\dagger H_k(t) \hat{\psi}_k,
\qquad 
\hat\psi_k = \begin{pmatrix} c^\dagger_k \\ c_{-k} \end{pmatrix}\\
\hat H_k(g_2) &=&(g_2 - \cos k) \Big( \lvert 1 \rangle_k \langle 1 \rvert - \lvert 0 \rangle_{k,-k} \langle 0 \rvert_{k,-k} \Big)
+ \sin k \Big( \lvert 1 \rangle_{k,-k} \langle 0 \rvert_{k,-k} + \lvert 0 \rangle_{k,-k} \langle 1 \rvert_{k,-k} \Big),\\
    \Pi_k(g) &=& \cos^2\frac{\theta_k(g)}{2} \, \lvert 0\rangle_{k,-k} \langle 0 \rvert_{k,-k}
- \frac{\sin \theta_k(g)}{2} \big( \lvert 0\rangle_{k,-k} \langle 1 \rvert_{k,-k} + \lvert 1\rangle_{k,-k} \langle 0 \rvert_{k,-k} \big)
\\
&&+ \sin^2\frac{\theta_k(g)}{2} \, \lvert 1\rangle_{k,-k} \langle 1 \rvert,\,
\theta_k(g) = \arctan \frac{\sin k}{g - \cos k},\nonumber\\
    \hat\rho_A &=& \prod_{k>0}\Pi_k(g_1),\quad\Pi_B = \prod_{k>0} \Pi_k(g_f).
\end{eqnarray}
The time-evolved projector becomes
\begin{eqnarray}
    U_k(t)&=&e^{-i \hat{H}_k(g_2) t}=\cos\big(\varepsilon_k(g_2) t\big) \, \mathbb{I}_k
- i \, \frac{\sin\big(\varepsilon_k(g_2) t\big)}{\varepsilon_k(g_2)}
H_k(g_2),\quad\varepsilon_k(g_2) = \sqrt{(g_2 - \cos k)^2 + \sin^2 k},\\
    \Pi_k(g_f,t) &=& U_k(t) \, \Pi_k(g_f) \, U_k^\dagger(t) \\
&=& u_{11,k}(t) \, \lvert 0\rangle_k \langle 0 \rvert
+ u_{12,k}(t) \big( \lvert 0\rangle_{k,-k} \langle 1 \rvert_{k,-k} + \lvert 1\rangle_{k,-k} \langle 0 \rvert_{k,-k} \big)
+ u_{22,k}(t) \, \lvert 1\rangle_{k,-k} \langle 1 \rvert_{k,-k},\nonumber\\
u_{11,k}(t) &=& \cos^2\frac{\theta_k(g_f)}{2} \cos^2(\varepsilon_k(g_2) t) + \sin^2\frac{\theta_k(g_f)}{2} \sin^2(\varepsilon_k(g_2) t)  - \frac{\sin\theta_k(g_f)}{2} \sin(2 \varepsilon_k(g_2) t),\\
u_{12,k}(t) &=& \frac{\sin\theta_k(g_f)}{2} \cos(2 \varepsilon_k(g_2) t) - \frac{1}{2} \sin(2 \varepsilon_k(g_2) t) \cos\theta_k(g_f),\\
u_{22,k}(t) &=& \sin^2\frac{\theta_k(g_f)}{2} \cos^2(\varepsilon_k(g_2) t) + \cos^2\frac{\theta_k(g_f)}{2} \sin^2(\varepsilon_k(g_2) t)  + \frac{\sin\theta_k(g_f)}{2} \sin(2 \varepsilon_k(g_2) t).
\end{eqnarray}
To this end, the transition probability to the $g_f$ target subspace generalizes the concept of thermalization after a quantum quench. First, we need the general 
\begin{eqnarray}
    P(B,t\vert A)&=&\tr[\hat \rho_A(t)\prod_k\Pi_k(g_f)]=\prod_k\tr[\Pi_k(g_1)\Pi_k(g_f,t)]\\
    &=&\prod_{k>0} \frac{1}{2} \Big[ 1 + \cos \theta_k(g_1) \cos \theta_k(g_f) - \sin \theta_k(g_1) \sin\theta_k(g_f) \, \cos(2 \varepsilon_k(g_2) t) \Big],\quad\theta_k=\arctan\left[\frac{\sin k}{g-\cos k}\right].\nonumber
\end{eqnarray}
This expression is essential to study the generalized final-state version of the Lochschmidt echo, which perfectly fits into the QTR framework by considering the final projector of $\hat\Pi_B=\mathbb I-\prod_k\hat\Pi_k(g_1)$. In this case, the transition probability and the corresponding QTR are given by
\begin{eqnarray}
    P(B,t\vert A)&=&1-\mathcal L(g_1,g_2)=1-\prod_k\frac{1}{2}\Big[ 1 + \cos^2 \theta_k(g_1)  - \sin^2 \theta_k(g_1) \, \cos(2 \varepsilon_k(g_2) t) \Big],\\
    k_{A\rightarrow B}(t)&=&-P(B,t\vert A)\sum_k\frac{\epsilon_k(g_2)\sin^2\theta_k(g_1)\sin\left(2\epsilon_k(g_2)t\right)}{ 1 + \cos^2 \theta_k(g_1)  - \sin^2 \theta_k(g_1) \, \cos(2 \varepsilon_k(g_2) t)}.
\end{eqnarray}
It is instructive to study their short and long-time behaviors. The short time limit is captured with $\cos(2\epsilon_k(g_2)t)\approx1-2\epsilon^2_k(g_2)t^2$,
\begin{eqnarray}
\mathcal L(t)&\approx&\mathcal \prod_{k>0}\cos^2\theta_k(g_1)\prod_{k>0}\left[1+\frac{\sin^2\theta_k(g_1)\epsilon^2_k(g_2) t^2}{\cos^2\theta_k(g_1)}\right]\equiv\mathcal L(0)\,\delta\mathcal L\\
&\approx&\mathcal L(0)\left[1+\sum_{k>0}\frac{\sin^2\theta_k(g_1)\epsilon^2_k(g_2) t^2}{\cos^2\theta_k(g_1)}\right]\approx\mathcal L(0)\left[1+\frac{L}{2\pi}\int_0^\pi\mathrm dk\frac{\sin^2\theta_k(g_1)\epsilon^2_k(g_2) t^2}{\cos^2\theta_k(g_1)}\right]\nonumber\\
&\approx&\mathcal L(0)\left[1+\frac{L}{4}\frac{(g_2-g_1)^2(2g^2_1-1)+g^2_1}{(1-g^2_1)^{3/2}}t^2\right].\nonumber
\end{eqnarray}

In the long-time limit, one can apply a stationary point approximation to the product as
\begin{eqnarray}
P(B,t| A)&=&1-\mathcal L,\\
\mathcal L &=& \prod_{k>0} \left[\frac{1+\cos^2\theta_k(g_1)}{2}-\frac{\sin^2\theta_k(g_1)}{2}\cos(2\epsilon_k(g_2)t)\right]\\
&=&\prod_{k>0}\frac{1+\cos^2\theta_k(g_1)}{2}\prod_{k>0}\left[1-\frac{\sin^2\theta_k(g_1)}{1+\cos^2\theta_k(g_1)}\cos(2\epsilon_k(g_2)t)\right]\nonumber\\
&=&\exp\left[\frac{L}{2\pi}\int_0^\pi\mathrm dk\,\log\left(\frac{\cos^2\theta_k(g_1)+1}{2}\right)\right]\exp\left[\frac{L}{2\pi}\int_0^\pi\mathrm dk\log(1-x(k)\cos(2\epsilon_k(g_2)t))\right]\nonumber\\
&\approx& F(g_1)\exp\Bigg[\frac{L}{2\pi}\int\mathrm dk\log\left[1-\frac{k^2}{2 (g_1-1)^2} \cos\Big( 2 |1-g_2| t + \frac{g_2}{|1-g_2|} t k^2 \Big)\right]\nonumber\\
&&+\frac{L}{2\pi}\int\mathrm dk\log\left[1-\frac{(\pi - k)^2}{2 (g_1+1)^2} \cos\Big( 2 (1+g_2) t - \frac{g_2}{1+g_2} t (\pi - k)^2 \Big)\right]\Bigg],\\
F(g_1)&\approx& \mathcal L(0),\,\text{if }g_1\gg1.\nonumber
\end{eqnarray}
Here, a stationary phase approximation has been employed with
$x(k)=\frac{\sin^2\theta_k(g_1)}{1+\cos^2\theta_k(g_1)}$ and the argument of the logarithm expanded up to second order around the stationary points, $k_0=0,\pi$. Now the logarithms can be expanded as the integral is restricted to small $k$-s close to the stationary point,
\begin{eqnarray}
    &&\int\mathrm dk\log\left[1-\frac{k^2}{2 (g_1-1)^2} \cos\Big( 2 |1-g_2| t + \frac{g_2}{|1-g_2|} t k^2 \Big)\right]\approx -\int\mathrm dk\frac{k^2}{2 (g_1-1)^2} \cos\Big( 2 |1-g_2| t + \frac{g_2}{|1-g_2|} t k^2 \Big)\nonumber\\
    &\approx&-\frac{1}{2(g_1-1)^2t^{3/2}}\int_0^\infty\mathrm du u^2\cos\left(2|1-g_2|t+\frac{g_2}{|1-g_2|}u^2\right)\nonumber\\
    &=&-\frac{\sqrt\pi}{8(g_1-1)^2}\left(\frac{|1-g_2|}{g_2}\right)^{3/2}\frac{\cos(2|1-g_2|t+3\pi/4)}{t^{3/2}}\\
    &&\int\mathrm dk\log\left[1-\frac{(\pi-k)^2}{2 (g_1+1)^2} \cos\Big( 2 (1+g_2) t - \frac{g_2}{1+g_2} t (\pi-k)^2 \Big)\right]
    \approx-\int\mathrm dk\frac{(\pi-k)^2}{2 (g_1+1)^2} \cos\Big( 2 (1+g_2) t - \frac{g_2}{1+g_2} t (\pi-k)^2 \Big)\nonumber\\
    &\approx&-\frac{1}{2(g_1+1)^2t^{3/2}}\int_0^\infty\mathrm du u^2\cos\left(2(1+g_2)t-\frac{g_2}{1+g_2)}u^2\right)=-\frac{\sqrt\pi}{8(g_1+1)^2}\left(\frac{1+g_2}{g_2}\right)^{3/2}\frac{\cos(2(1+g_2)t-3\pi/4)}{t^{3/2}}.
\end{eqnarray}
Thus, the final long-time behavior is given by,
\begin{eqnarray}
    &&P(B,t\vert A)\approx 1-F(g_1)\\
    &&\times\exp\left[-\frac{\sqrt\pi}{8(g_1+1)^2}\left(\frac{1+g_2}{g_2}\right)^{3/2}\frac{\cos(2(1+g_2)t-3\pi/4)}{t^{3/2}}-\frac{\sqrt\pi}{8(g_1-1)^2}\left(\frac{|1-g_2|}{g_2}\right)^{3/2}\frac{\cos(2|1-g_2|t+3\pi/4)}{t^{3/2}}\right]\nonumber\\
    &&\approx1-F(g_1)\nonumber\\
    &&\times\left[1-\frac{\sqrt\pi}{8(g_1+1)^2}\left(\frac{1+g_2}{g_2}\right)^{3/2}\frac{\cos(2(1+g_2)t-3\pi/4)}{t^{3/2}}-\frac{\sqrt\pi}{8(g_1-1)^2}\left(\frac{|1-g_2|}{g_2}\right)^{3/2}\frac{\cos(2|1-g_2|t+3\pi/4)}{t^{3/2}}\right].\nonumber
\end{eqnarray}
The corresponding QTRs can be obtained by taking the time derivatives in the given limits. For short times, it is relatively simple,
\begin{eqnarray}
    k_{A\rightarrow B}(t)&\approx&-2\prod_{k>0}\cos^2\theta_k(g_1)\frac{L}{2}\frac{(g_2-g_1)^2(2g^2_1-1)+g^2_1}{(1-g^2_1)^{3/2}}t\\
    &\approx& -2\exp\left[\frac{L}{2\pi}\int_0^\pi\mathrm dk\log\left[\frac{(g_1-\cos k)^2}{1+g^2_1-2g_1\cos k}\right]\right]\frac{L}{2}\frac{(g_2-g_1)^2(2g^2_1-1)+g^2_1}{(1-g^2_1)^{3/2}}t\nonumber\\
    &=&-2\exp\left[L\log\left[\frac{g_1+\sqrt{g^2_1-1}}{2g_1}\right]\right]\frac{L}{2}\frac{(g_2-g_1)^2(2g^2_1-1)+g^2_1}{(1-g^2_1)^{3/2}}t.\nonumber
\end{eqnarray}
For long times, one needs to be more cautious, as only the derivatives of the cosines of the exponentials survive in that limit
\begin{eqnarray}
    &&k_{A\rightarrow B}(t)\approx F(g_1)\\
    &&\times\left[\frac{\sqrt\pi(1+g_2)}{4(g_1+1)^2}\left(\frac{(1+g_2)}{g_2}\right)^{3/2}\sin(2(1+g_2)t-3\pi/4)+\frac{\sqrt\pi|1-g_2|}{4(g_1-1)^2}\left(\frac{|1-g_2|}{g_2}\right)^{3/2}\sin(2|1-g_2|t+3\pi/4)\right]t^{-3/2}.\nonumber
\end{eqnarray}
To this end, we also compute the corresponding QSL based on the transition probability. The Bures angle in the short and long time limits is given by
\begin{eqnarray}
    &&\cos^{-1}\sqrt{P(B,t\vert A)}\approx\frac{\pi}{2}-\sqrt{\frac{L}{2}\frac{(g_2-g_1)^2(2g^2_1-1)+g^2_1}{(1-g^2_1)^{3/2}}}\exp\left[L\log\left(\frac{1+\sqrt{1-g^2_1}}{4}\right)\right]t,\\
    &&\cos^{-1}\sqrt{P(B,t\vert A)}\approx \cos^{-1}\sqrt{1-F(g_1)}-\frac{F^{3/2}(g_1)}{2\sqrt{1-F(g_1)}}\\
    &&\times\left[\frac{\sqrt\pi}{8(g_1+1)^2}\left(\frac{1+g_2}{g_2}\right)^{3/2}\frac{\cos(2(1+g_2)t-3\pi/4)}{t^{3/2}}+\frac{\sqrt\pi}{8(g_1-1)^2}\left(\frac{|1-g_2|}{g_2}\right)^{3/2}\frac{\cos(2|1-g_2|t+3\pi/4)}{t^{3/2}}\right].\nonumber
\end{eqnarray}
For the QSL, the energy variance is given by
\begin{eqnarray}
    \Delta^2_{\hat\rho_A}\hat H(g_2)&\equiv&\Delta^2_{\hat\Pi(g_1)}\hat H(g_2)=\sum_{k>0}\Delta^2_{\hat\Pi_k(g_1)}\hat H_k(g_2)\\
    &=&\sum_{k>0}\langle\hat H^2_k(g_2)\rangle_{\hat\Pi_k(g_1)}-\langle\hat H_k(g_2)\rangle^2_{\hat\Pi_k(g_1)}=\sum_{k>0}\epsilon^2_k(g_2)-\frac{((g_2-\cos k)(g_1-\cos k)-\sin^2k)^2}{\epsilon^2_k(g_1)}\nonumber\\
    &=&\sum_{k>0}\frac{(g_2-g_1)^2\sin^2 k}{\epsilon^2_k(g_1)}\approx\frac{L}{2\pi}\int_0^\pi\mathrm dk\,\frac{(g_2-g_1)^2\sin^2 k}{\epsilon^2_k(g_1)}\nonumber\\
    &=&\frac{L}{2\pi}(g_2-g_1)^2\int_0^\pi\mathrm dk\,\frac{\sin^2 k}{g^2_1-2g_1\cos k+1}=\frac{L}{4}\frac{(g_2-g_1)^2}{g^2_1}.\nonumber
\end{eqnarray}

Next, we also establish the relation of the QTRs with the correlation functions following a similar quantum quench starting from the ferromagnetic ground state, $\hat\rho_A=\hat\Pi(0)$ evolved by $\hat H(g)$. To this end, we choose the initial density matrix of
\begin{eqnarray}
    \hat\rho_A&=&\prod_{k>0}\hat\Pi_k(0),\\
    \hat\Pi_k(0)&=&\sin^2\frac{k}{2}\lvert0\rangle_k\lvert0\rangle_{-k}\langle0\rvert_{-k}\langle0\rvert_k-\frac{\sin k}{2}\Big(\lvert1\rangle_k\lvert1\rangle_{-k}\langle0\rvert_{-k}\langle0\rvert_k+\lvert0\rangle_k\lvert0\rangle_{-k}\langle1\rvert_{-k}\langle1\rvert_k\Big)+\cos^2\frac{k}{2}\lvert1\rangle_k\lvert1\rangle_{-k}\langle1\rvert_{-k}\langle1\rvert_k.\qquad
\end{eqnarray}
Choosing the target subspace $\hat\Pi_B=(1-\hat\sigma^z_k)/2$, the initial orthogonality condition as $\tr[\hat\rho_A\hat\Pi_B]=0$ is satisfied, and one also recovers the one-point correlation function by the transition probability as
\begin{eqnarray}
    &&P(B,t\vert A)=\tr[\hat\rho_A(t)\hat\Pi_B]=\frac{1-\langle\sigma^z_j(t)\rangle}{2},\,\langle\sigma^z_j(t)\rangle\propto\exp\left[t\int_0^\pi\mathrm dk\,\frac{g\sin k}{\sqrt{1+g^2-2g\cos k}}\log\left[\cos \Delta_k(g)\right]\right],\\
    &&\cos \Delta_k(g)=\frac{1-g\cos k}{4\sqrt{1+g^2-2g\cos k}},\nonumber
\end{eqnarray}
given in Ref.~\cite{Calabrese_PRL2011, Calabrese_2012Review, Calabrese_2012_Review2}.
The corresponding QTR, thus, reveals further, yet unknown characteristics of the rate of change of the correlation functions
\begin{eqnarray}
    k_{A\rightarrow B}(t)=-i\tr[\hat\rho_A[\hat\Pi_B(t),\hat H(g)]].
\end{eqnarray}
For the evaluation, one needs to transfer the time evolution to the density matrix, by which the commutator becomes
\begin{eqnarray}
    &&\hat\rho_A(t)=e^{it\hat H_k(g)}\rho_A e^{-it\hat H_k(g_2)},\,[\hat\Pi_B,\hat H]=-i\frac{g}{2}\sigma^y_j,\\
    &&k_{A\rightarrow B}(t)=-\frac{g}{2}\tr[\hat\rho_A\hat\sigma^y_j(t)],
\end{eqnarray}
which exhibits a similar structure as the one-point correlation function $\langle\hat\sigma^z_j(t)\rangle$, which can only be captured by a complicated Pfaffian, and exact analytics is only possible in the long-time limit. Remarkably, the QTR formulation can achieve a simple bound in terms of the transition probability without requiring long, complex calculations
\begin{eqnarray}
    \lvert k_{A\rightarrow B}(t)\rvert&\propto&\tr[\hat\rho_A\left[\hat\sigma^z_j(t),\hat H\right]]\leq2\Delta_{\hat\rho_A}\hat H(g)\sqrt{\langle\hat\sigma^z_j(t)\rangle}\sqrt{1-\langle\hat\sigma^z_j(t)\rangle}\\
    &\propto&\sqrt Lg\sqrt Lg\exp\left[\frac{t}{2}\int_0^\pi\frac{\mathrm dk}{\pi}\,\epsilon'_k(g)\log\left[\Delta_k(g)\right]\right],\epsilon'_k(g)=\frac{g\sin k}{\sqrt{1+g^2-2g\cos k}},\nonumber
\end{eqnarray}
in the leading order up to exponential accuracy.
This bound on the associated QTR also allows for characterizing the QSL for the change of the one-point function, i.e., the minimal time, under which local magnetization can change,
\begin{eqnarray}
\tau_z\equiv\frac{\arccos{\sqrt{\frac{\lvert\langle\hat\sigma^z_j(t)\rangle\rvert}{2}}}}{\sqrt Lg},
\end{eqnarray}
as the rank of the target subspace is $2$ identified with the general parameter $d_B=2$ in the general setting, and we have used the results for the energy variance.
The construction naturally follows for the two-point correlation function as well~\cite{Calabrese_PRL2011}, by choosing $\hat\Pi_B=(1-\hat\sigma^z_j\hat\sigma^z_{j+l})/2$. The corresponding transition probability is given by
\begin{eqnarray}
    &&P(B,t\vert A)=\frac{1-\langle\hat\sigma^z_j(t)\hat\sigma^z_{j+l}(t)\rangle}{2},\quad\langle\hat\sigma^z_j(t)\hat\sigma^z_{j+l}(t)\rangle\propto \exp\left[l\int_0^\pi\frac{\mathrm dk}{\pi}\log\left[\cos\Delta_k(g)\right]\Theta\left(2\epsilon'_k(g)t-l\right)\right]\nonumber\\
    &&\times\exp\left[2t\int_0^\pi\frac{\mathrm dk}{\pi}\epsilon'_k(g)\log\left[\cos\Delta_k(g)\right]\Theta\left(2\epsilon'_k(g)t-l\right)\right],
\end{eqnarray}
From here, the associated transition rate can also be upper-bounded for long times by similar means,
\begin{eqnarray}
    \lvert k_{A\rightarrow B}(t)\rvert&=&\lvert\tr[\hat\rho_A[\hat\sigma^z_j\hat\sigma^z_{j+l},\hat H]]\rvert\leq 2\Delta_{\hat\rho_A} \hat H\sqrt{\langle\hat\sigma^z_j(t)\hat\sigma^z_{j+l}(t)\rangle}\sqrt{1-\langle\hat\sigma^z_j(t)\hat\sigma^z_{j+l}(t)\rangle}\\
    &\propto& \sqrt{Lg}\exp\left[\frac{l}{2}\int_0^\pi\frac{\mathrm dk}{\pi}\log\left[\cos\Delta_k(g)\right]\Theta\left(2\epsilon'_k(g)t-l\right)\right]\exp\left[t\int_0^\pi\frac{\mathrm dk}{\pi}\epsilon'_k(g)\log\left[\cos\Delta_k(g)\right]\Theta\left(2\epsilon'_k(g)t-l\right)\right].\nonumber
\end{eqnarray}
Finally, the bound on the rate of change of $\langle\hat\sigma^z_j(t)\hat\sigma^z_{j+l}(t)\rangle$ again allows us to characterize the maximum rate, i.e., the minimum time for a substantial change at which the two-point correlation function can vary,
\begin{eqnarray}
    \tau_{zz}\equiv \frac{\arccos\sqrt{\frac{\lvert\langle\hat\sigma^z_j(t)\hat\sigma^z_{j+l}(t)\rangle\rvert}{4}}}{\sqrt{L}g}.
\end{eqnarray}

\end{document}